\documentclass[PRL,aps,twocolumn,superscriptaddress,10pt]{revtex4-1}

\usepackage{multirow}
\usepackage{textgreek}
\usepackage [english]{babel}
\usepackage [autostyle, english = american]{csquotes}
\MakeOuterQuote{"}

\usepackage{graphicx}
\usepackage{epstopdf}
\usepackage{color}
\usepackage{amsmath,amssymb}
\begin{document}

\title{Giant Domain Walls and Intrinsic Heterogeneity in 214 Cuprate Superconductors}

\author{Evie Ladbrook} 
\affiliation{Department of Chemistry, University of Warwick, Gibbet Hill, Coventry, CV4 7AL, United Kingdom}

\author{Jon P. Wright} 
\affiliation{European Synchrotron Radiation Facility, 38000 Grenoble, France}

\author{Mark S. Senn} 
\email{m.senn@warwick.ac.uk}
\affiliation{Department of Chemistry, University of Warwick, Gibbet Hill, Coventry, CV4 7AL, United Kingdom}

\date{\today}%

\begin{abstract}

Structural phase transitions generate complex microstructures that often govern material functionality, yet directly resolving their three-dimensional organisation in bulk samples remains challenging. Here we employ scanning three-dimensional X-ray diffraction to resolve the bulk microstructure of La$_{1.675}$Eu$_{0.2}$Sr$_{0.125}$CuO$_{4}$, a prototypical 1/8-doped cuprate in which structural and electronic heterogeneity is well established. We reveal remarkably broad tetragonal-like domain wall regions within the nominally orthorhombic crystal structure, and, upon cooling to 100 K, a fine microstructure of orthorhombic-like stripes embedded within the tetragonal matrix that has significant consequences for interpreting the interplay between structural and electronic heterogeneity in this class of materials. More broadly, this work establishes 3DXRD as a powerful approach for resolving bulk microstructures and understanding their role in emergent functionality.

\end{abstract}

\maketitle

The physical properties of materials are often governed by their microstructure. In ferroelastic and martensitic systems, the accommodation of strain during symmetry-lowering phase transitions generates complex microstructures spanning multiple length scales. Interfaces formed during these transitions can themselves become active elements, hosting properties absent in the surrounding bulk, including enhanced conductivity \cite{seidel_conduction_2009}, polarity \cite{zubko_strain-gradient-induced_2007} and ferroelectricity \cite{tao_ferroelectric_2024}, motivating intense interest in understanding and controlling microstructure. Despite its importance, directly visualising bulk microstructure over the relevant length scales remains challenging. Electron microscopy and scanning probe techniques provide high spatial resolution but are generally restricted to surfaces or thin samples, while many bulk probes average over spatial correlations. As a result, establishing how local structural distortions organise into extended three-dimensional microstructures remains an outstanding challenge across a wide range of functional materials.

The consequences of such microstructural heterogeneity are particularly evident in correlated electron materials, where strong coupling between strain, charge and spin degrees of freedom can amplify the effects of local structural variations. Inhomogeneities, whether structural, electronic or chemical, can drive phase separation on the nanometre to micrometre scale, with significant consequences for their properties \cite{dagotto_complexity_2005, campi_inhomogeneity_2015}. This is notably observed in the manganites, where the colossal magnetoresistance is believed to be linked to the percolation of a ferromagnetic conducting phase within a charge-ordered antiferromagnetic insulating matrix \cite{uehara_percolative_1999}. 

High-temperature superconducting cuprates provide another important example in which structural and electronic inhomogeneity are ubiquitous. Their layered perovskite framework supports cooperative rotations and tilts of the CuO$_6$ octahedra, giving rise to structural phase transitions that are strongly coupled to charge, spin and superconducting degrees of freedom. This interplay manifests in phenomena including charge stripes \cite{tranquada_evidence_1995}, nematicity \cite{hinkov_electronic_2008} and filamentary superconductivity \cite{phillips_filamentary_1997}. A wide range of experimental techniques has revealed evidence for structural and electronic inhomogeneity across multiple length scales. Local structural probes, including pair distribution function analysis \cite{egami_lattice_1994} and extended X-ray absorption fine structure measurements \cite{saini_evidence_2001}, reveal deviations from the average crystallographic structure, while scanning tunnelling microscopy has demonstrated substantial spatial variations in electronic properties \cite{pan_microscopic_2001,pasupathy_electronic_2008}. Collectively, these studies establish that inhomogeneity is an intrinsic feature of the cuprates. Despite this extensive body of work, the three-dimensional bulk organisation of these structural distortions remains unresolved. Bulk probes can reveal local distortions but not their spatial arrangement, while surface-sensitive techniques cannot access the bulk microstructure. Consequently, the interplay between structural and electronic heterogeneity, and the effect this has on bulk properties remains poorly understood. 

La$_{1.675}$Eu$_{0.2}$Sr$_{0.125}$CuO$_{4}$ (LESCO), a prototypical 1/8-doped cuprate \cite{moodenbaugh_superconducting_1988, fink_phase_2011}, provides an ideal model system to address this. LESCO exhibits signatures of structural coexistence between the low-temperature orthorhombic (LTO) and low-temperature tetragonal (LTT) phases over a broad temperature range \cite{tidey_pronounced_2022}, reminiscent of electronic phase separation observed in other highly correlated systems. While these signatures are also evident in several 1/8-doped cuprates, and notably absent in the isostructural band insulator La$_{2}$MgO$_{4}$ \cite{tidey_structural_2022}, prevailing interpretations and theories often assume a homogeneous structural state underpins both superconducting and charge density wave (CDW) order.

\begin{figure*}
\includegraphics[width=\textwidth]{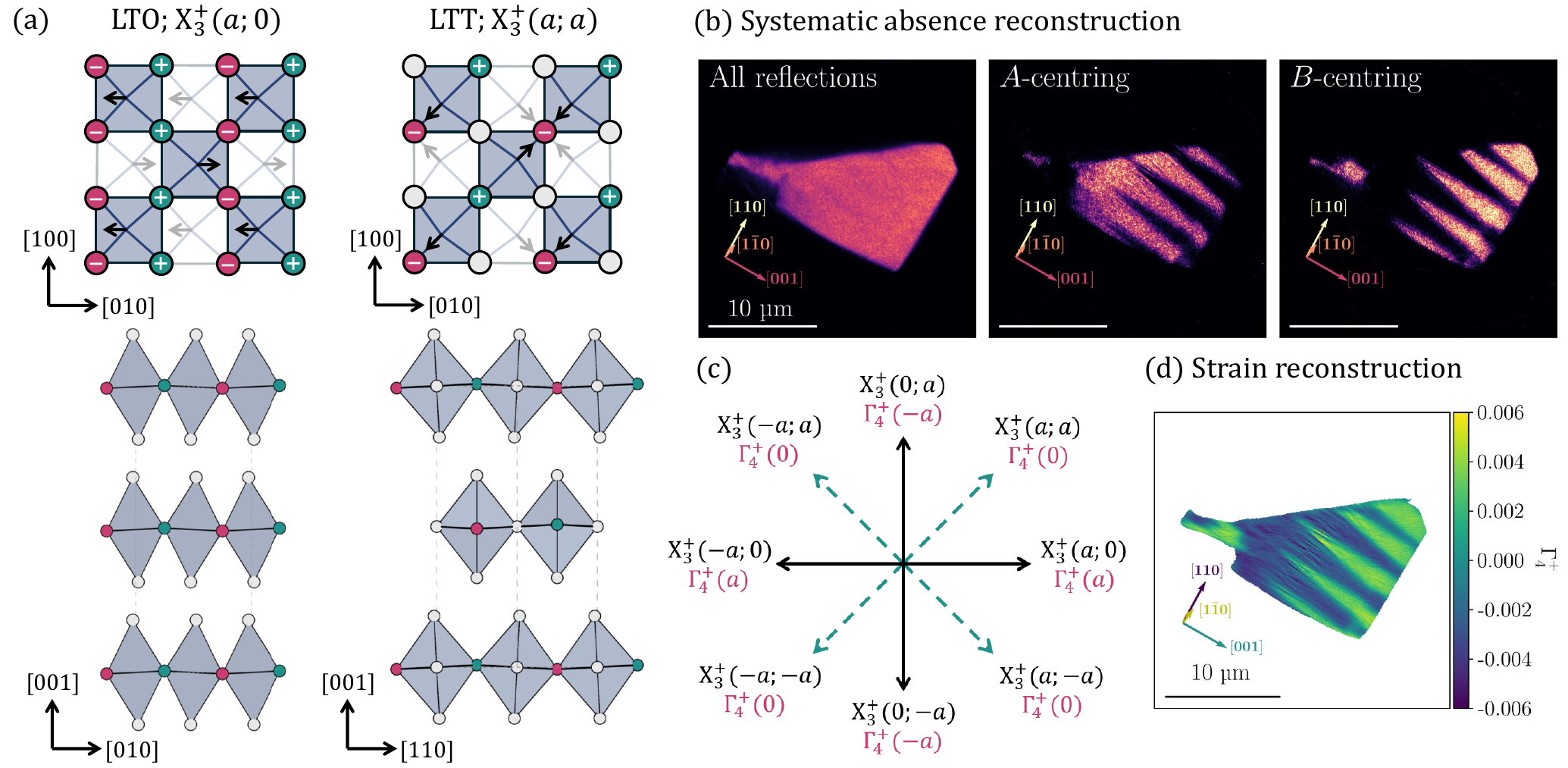}
\caption{(a) Representation of the LTO (left) and LTT (right) structures, with $A$-site cation omitted for clarity. Oxygens are shown in teal and magenta indicating whether they are above and below the CuO$_2$ planes, respectively. In LTO, the sense of the distortion propagates along [010], with all layers aligned. In LTT, the sense of the distortion propagates along alternating diagonals, with a 90$^\circ$ rotation, from [110] to [1$\Bar{1}$0], between adjacent layers. (b) 300 K reconstruction of the sample from all reflection and filtered by $A$- and $B$-centred systematic absences, revealing twin domains. (c) Order parameter map indicating X$_{3}^{+}$ propagation direction with black as LTO and teal as LTT. (d) 300 K strain reconstruction based on $\Gamma_{4}^{+}$ showing strong correlations with systematic absence map. Axes are given with respect to the LTO/LTT basis.}
\label{1}
\end{figure*}

At room temperature, LESCO adopts the LTO ($Bmab$) phase, characterised by rotations of the CuO$_6$ octahedra about the [100] axis, as shown in Fig.~\ref{1}a. This distortion transforms as irreducible representation X$_{3}^{+}$($a;0$), with respect to the high-temperature tetragonal (HTT, $I4/mmm$, $F4/mmm$ in the basis of LTO) phase. Upon cooling to approximately 160 K, LESCO undergoes a first-order phase transition from LTO to the LTT ($P4_{2}/ncm$) phase, in which the octahedra rotate about [110] and [1$\Bar{1}$0], with alternating layers rotating in opposite senses. This is described by X$_{3}^{+}$($a;a$). The notation used here throughout derives from representation theory, also termed symmetry-mode analysis, and is used to describe the distorted structures ($Bmab$ or $P4_{2}/ncm$) in terms of the high-symmetry parent structure ($I4/mmm$) plus a combination of symmetry-adapted distortions. In this way, we can establish a direct microscopic description of the order parameters. This information can be extracted using software such as ISODISTORT \cite{isotropy, campbell_b_j_isodisplace_2006}. The absence of a direct group-subgroup relationship between LTO and LTT necessitates a region of phase coexistence. However, the persistence of this down to at least as low as 10 K in LESCO and isostructural La$_{1.875}$Ba$_{0.125}$CuO$_{4}$ (LBCO) suggests a coupling to electronic phase separation related to an intrinsic competition between superconductivity and CDW order at this doping level. While CDW order is believed to suppress bulk superconductivity \cite{tranquada_evidence_1995}, in-plane superconducting correlations may persist \cite{hanaguri_checkerboard_2004, berg_dynamical_2007}. In the LTO phase, charge stripe alignment across layers enables bulk superconducting coherence, but in the LTT phase, the octahedral rotations pin charge stripes and induce a 90$^\circ$ rotation between adjacent layers, frustrating interlayer Josephson coupling \cite{fujita_stripe_2004, sears_structure_2023}. Similar frustrations may occur at structural domain walls, where the sense of the octahedral rotations reverses, inducing corresponding charge stripe rotations. Therefore, a detailed understanding of the microstructure at the domain-level appears essential for accurately measuring and controlling superconductivity.

To directly resolve this complex microstructure, we used scanning three-dimensional X-ray diffraction \cite{poulsen_applications_1997, wright_new_2020} at ID11 at the ESRF \cite{ESRF_data}, allowing spatially resolved characterisation of LTO- and LTT-like structural domains in the bulk from 300 to 100 K. This method contrasts techniques like transmission electron microscopy (TEM), which typically only provide surface-sensitive information, as it instead offers a cross section through the bulk. Similarly to conventional single crystal X-ray diffraction, the full diffraction signal is collected. This contrasts with many alternative X-ray and electron imaging techniques that measure from only specific reflections. In the resulting dataset, the position of each reflection is characterised by its scattering angle (2$\theta$) and azimuthal angle ($\eta$), which is then related to its spatial origin within the sample through the rotation angle ($\omega$) and sample translation ($y$). The additional measurement of the sample translation distinguishes 3DXRD from conventional single crystal diffraction, where only angular dimensions are recorded.  Using the full diffraction signal indexed on a 5.35~$\times$~5.35~$\times$~13.2 \AA~unit cell, the shape of the slice through the crystal can be tomographically reconstructed (Fig.~\ref{1}b). A more insightful view of the underlying twin domain structure emerges by performing the reconstruction using only a subset of reflections. The two orthorhombic twin domains, related by a fourfold rotation about the $c$-axis, have different systematic absence conditions relative to the lattice used to index all observed reflections. They correspond to either an $A$- or $B$-centred lattice, characterised by general reflection conditions $k + l =$ even or $h + l =$ even, respectively. This distinction arises from the orientation of the octahedral tilting, as illustrated in Fig~{1}c. Reflections where $h,k,l =$ even are excluded in filtering. This clearly resolves a micron-scale domain structure composed of interlocking $A$- or $B$-centred regions, orientated parallel to [001] and perpendicular to [110] and [1$\bar1$0], likely forming continuous sheets in the out-of-plane direction. The [001] direction is approximately 2$^\circ$ out of the plane of the image which may account for the triangular shape of the domains if the sheets are not of a constant thickness throughout the crystal. 

As the LTT structure emerging at lower temperature lacks general systematic absence conditions to differentiate it from the LTO domains, contrast was instead achieved through variation in the ferroelastic strain, $\Gamma_{4}^{+}$, that acts as a secondary order parameter of the octahedral tilting, X$_{3}^{+}$. Iterative reconstructions based on the spatial distribution of the local average unit cell parameters \cite{henningsson_efficient_2023} (further details provided in the Supplemental Material \cite{Supplemental_Material}) recovered Fig.~\ref{1}d. The strong correlation between the systematic absence reconstruction (from reflection intensities, Fig.~\ref{1}b) and the strain reconstruction (from peak positions, Fig.~\ref{1}d) provides robust validation of our approach. Given its superior resolution (10$^{-4}$ \cite{borbely_calibration_2014}) relative to systematic intensity errors, we focus our subsequent analysis on strain reconstructions. This provides a level of detail generally hard to achieve with techniques like TEM over such large length scales.

\begin{figure}[h]
\includegraphics[width=0.5\textwidth]{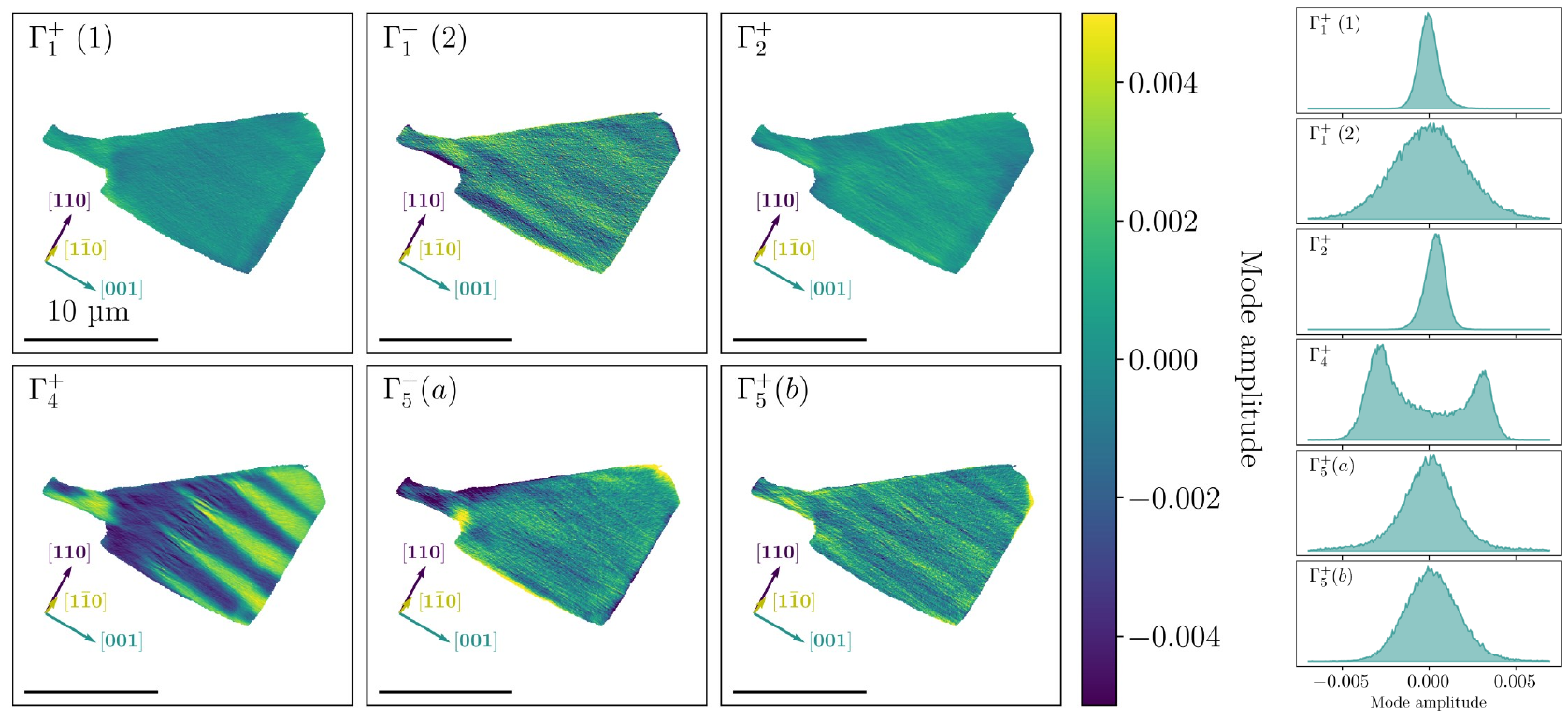}
\caption{Spatially resolved maps and associated histograms of symmetry-adapted strain modes at 300 K. Note the bimodal distribution of $\Gamma_{4}^{+}$, unique in capturing the twin domain structure. $\Gamma_{1}^{+}(1)$ corresponds to an in-plane symmetric expansion/contraction, while $\Gamma_{1}^{+}(2)$ is a uniform strain along $c$. $\Gamma_{2}^{+}$ is an in-plane shear strain. $\Gamma_{4}^{+}$ is the orthorhombic strain. $\Gamma_{5}^{+}(a)$ and $\Gamma_{5}^{+}(b)$ describe shear strains between the $ab$ plane and $c$ axis.}
\label{2}
\end{figure}

Fig.~\ref{2} shows the spatial variation of the symmetry adapted strain with respect to the $I4/mmm$ aristotype and associated histograms at 300 K.

Among these, $\Gamma_{4}^{+}$ is unique in exhibiting a pronounced, stripe-like pattern, directly aligned with the orthorhombic twin structure. In contrast, the other strain modes show only subtle spatial deviations, most notably in $\Gamma_{1}^{+}$(2), which corresponds to an elongation of \textit{c}, and $\Gamma_{5}^{+}$(b), associated with a shearing between \textit{c} and the \textit{ab} plane. We note that the small spatial variation in $\Gamma_{5}^{+}(b)$, most likely arising from strain localised at the domain walls, may provide an explanation for the apparent subtle monoclinic distortion reported in LSCO and LBCO \cite{frison_crystal_2022, hu_charge_2025}. This distribution is broader than some of the other minor modes (Fig. S3), but the magnitude of these spatial variations remains roughly three times smaller than that of $\Gamma_{4}^{+}$, the dominant orthorhombic strain. The corresponding histograms reinforce these observations: $\Gamma_{4}^{+}$ exhibits a clear bimodal distribution reflecting the alternating strain in the twin domains, with negative and positive peaks corresponding to the $A$- and $B$-domains, and a central region showing the gradual transition across the domain wall. By comparison, the other strain modes show uniform distributions centred around zero, consistent with the stable or weakly varying distortions. These can be well modelled by a single pseudo-Voigt distribution and would, for example, give rise to anisotropic peak broadening due to sample microstrain in powder XRD. This symmetry-based decomposition reveals that $\Gamma_{4}^{+}$ uniquely captures the structural signature of the octahedral tilting. 

\begin{figure}[h]
\includegraphics[width=0.5\textwidth]{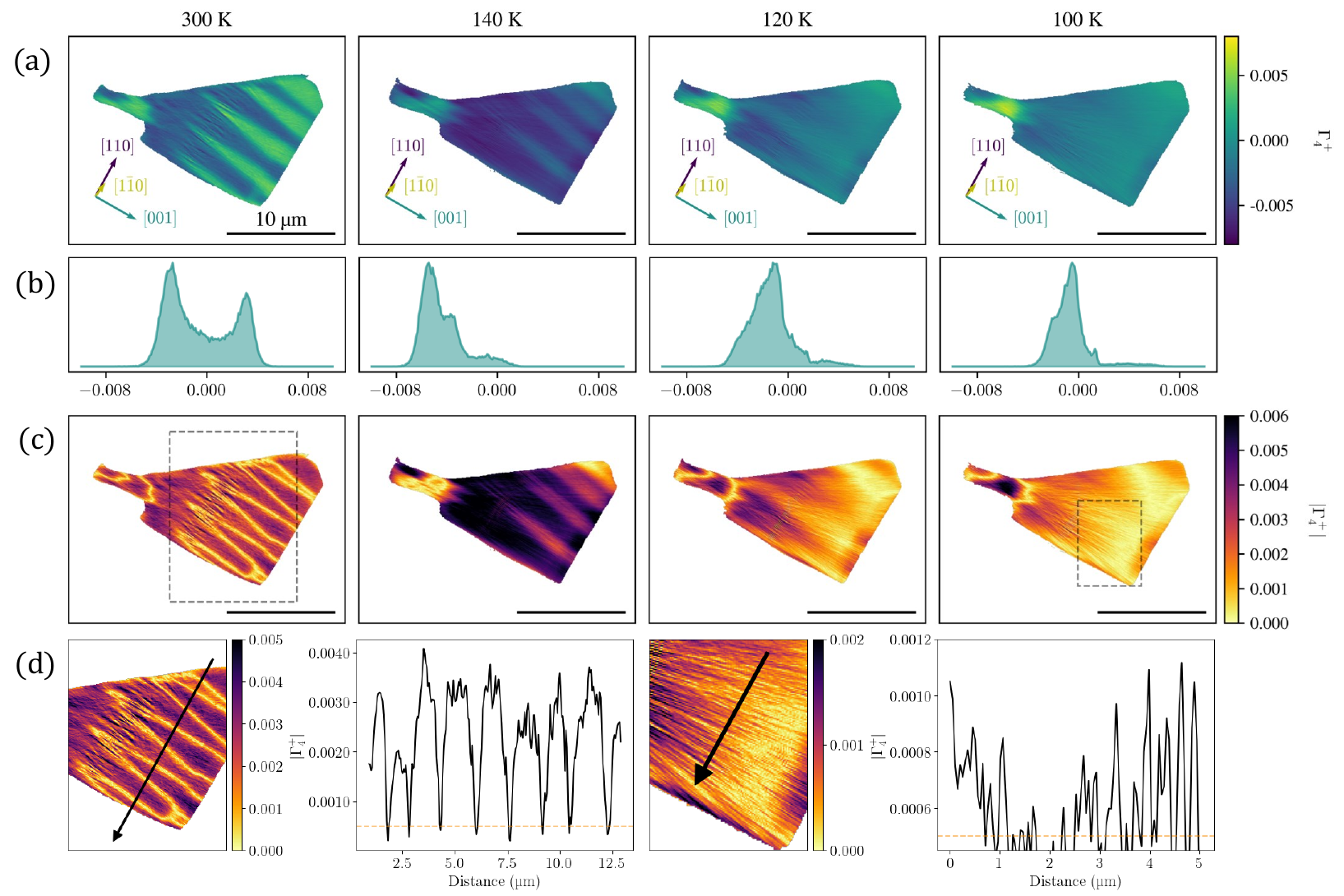}
\caption{(a) Spatially resolved maps of $\Gamma_{4}^{+}$ at 300 K, 140 K, 120 K and 100 K, illustrating the temperature evolution of the structural domains. (b) Histograms of $\Gamma_{4}^{+}$ at corresponding temperatures, showing the transition from bimodal at 300 K to a more homogenous but still asymmetric distribution at 100 K. (c) Maps of the absolute strain magnitude $|\Gamma_{4}^{+}|$ with (d) showing enhanced maps and line profiles, revealing broad LTT-like features within the LTO phase at 300 K and subtle LTO-like filaments within the LTT phase at 100 K. The spatial resolution of our reconstructions is at least 70 nm, based on the determined strain resolution of the experiment of 2$~\times~10^{-4}$, see the Supplemental Material \cite{Supplemental_Material}.}
\label{3}
\end{figure}

The evolution of $\Gamma_{4}^{+}$ upon cooling (Fig.~\ref{3}a-b) provides critical insight into the LTO/LTT transition. From 300 to 140 K, the $A$-centred LTO domain gradually dominates to accommodate the growing ferroelastic strain, leading to a marked asymmetry in the $\Gamma_{4}^{+}$ distribution (Fig.~\ref{3}b). However, before the minority domain is fully consumed, the increasing magnitude of X$_{3}^{+}$, inherently coupled to $\Gamma_{4}^{+}$, triggers the first-order transition to LTT \cite{tidey_structural_2022}. This behaviour, occurring near the critical octahedral tilting amplitude for $T_C$, is not a coincidence but instead suggests a structural basis underlying the suppression of superconductivity in the composition-doping phase diagram of this material class \cite{tidey_pronounced_2022}. At 120 and 100 K, the overall micro-scale variation in $\Gamma_{4}^{+}$ appears to be lost, consistent with a more homogenous LTT structure. However, analysis of the absolute strain magnitude $|\Gamma_{4}^{+}|$ (Fig.~\ref{3}c) uncovers crucial sub-micron variation at all temperatures, reminiscent of tweed domain structures \cite{salje_microstructures_1991}.

At 300 K, within the nominally LTO phase, our $|\Gamma_{4}^{+}|$ strain map reveals unusually wide, domain wall-like regions with near-zero strain. These features are not atomic-scale defects, but extended regions approximately 150 nm thick and separated by 0.5 -- 2 \textmu m (see Fig.~\ref{s6}/Table~\ref{tabs1} \cite{Supplemental_Material}). This length scale is orders of magnitude larger than typical ferroelastic or ferroelectric domain walls. Consequently, these structures challenge the more conventional understanding of what we would view as a domain wall or as microscopic coexisting phases with distinct symmetry. Instead, the observed walls are effectively extended LTT-like regions coexisting within an LTO matrix \cite{croft_charge_2014, wang_high-temperature_2020}. Their substantial thickness could even give rise to coherent scattering signals observable as distinct phases in conventional XRD, offering a natural explanation for the phase coexistence reported in LESCO and related cuprates \cite{tidey_pronounced_2022, axe_structural_1989, crawford_lattice_1991}. However, we note that the ability to resolve distinct structures, does not just depend on the coherence of the order parameters within the material, but also the resolution of the instrument and coherence of the radiation source measuring them.  

Considering Fig.~\ref{1}c, we construct a microscopic model for the evolution of the tilt order parameter across these domains. The strain profile shows a smooth transition from positive to negative $\Gamma_{4}^{+}$ through extended regions where $\Gamma_{4}^{+}$ approaches zero. This behaviour is consistent with a continuous rotation of the tilt, for example a sequence such as X$_{3}^{+}$($a;0$)$|$X$_{3}^{+}$($a;a$)$|$X$_{3}^{+}$($0;a$). Between the LTO domains and the LTT domain wall centre, the octahedral tilt gradually changes direction, potentially passing through a region that can be described with $Pccn$ symmetry, which is a common subgroup of LTO and LTT. A microscopic (crystallographic) picture for these possible domain evolutions is given in the Supplemental Material \cite{Supplemental_Material} (Fig.~\ref{dw}). The presence of local tetragonal character is consistent with early structural studies on related cuprates \cite{zhu_tetragonal-orthorhombic_1994, horibe_direct_2000, braden_characterization_1992, horibe_microstructure_1997} which suggest that ideally sharp orthorhombic twin walls are unlikely because they would require the CuO$_6$ octahedra to tilt abruptly in orthogonal directions, resulting in a high interfacial energy. Instead, extended boundaries are expected, in which the tilt pattern gradually transitions from orthorhombic to tetragonal at the wall centre. TEM measurements on LBCO estimated the width of these tetragonal regions to be around 100 \AA~, while neutron diffraction indicated larger values of a few hundred \AA~based on superstructure peak broadening \cite{zhu_tetragonal-orthorhombic_1994}. Consistently, Landau-type free-energy models for LBCO predict that the central region of such walls should adopt LTT-like character rather than revert to the HTT phase \cite{cai_theory_1994}. 

Anomalously wide and structurally coherent walls have been increasingly recognised in ferroic materials, but not widely discussed in superconducting materials. In Ca$_3$Ti$_2$O$_7$, infrared nano-spectroscopy revealed 60 -- 100 nm wide ferroelastic walls, attributed to a gradual rotation of the octahedral tilt and rotation order parameters \cite{smith_infrared_2019}. Similarly, N\'eel-like ferroelectric domain walls, where the polarisation rotates rather than vanishes, have been predicted and observed in systems such as BaTiO$_3$ \cite{yudin_anomalously_2015, lee_hidden_2017}. These examples support a broader understanding of wide domain boundaries as intrinsic structural features that can serve as functionally active elements \cite{xia_giant_2024, aird_sheet_1998}.

At 100 K, within the CDW phase, the microstructure evolves markedly. Deviations from ideal tetragonal symmetry emerge as fine stripe-like orthorhombic features on the 100 -- 200 nm scale, repeating with a similar periodicity. These may be interpreted as nanoscale inclusions of the LTO phase embedded within the LTT matrix, likely forming to minimise the elastic energy associated with antiphase boundaries related to the octahedral tilt orientation (X$_{3}^{+}$($a;a$)/X$_{3}^{+}$($-a;a$)). While similar behaviour has been reported previously in LBCO, where electron microscopy revealed a fine mixture of domains with orthorhombic and tetragonal character with sizes of $\geq$ 300 \AA~that persisted well below the LTO/LTT transition to 10 K \cite{chen_low_1991, chen_micro-twin_1991}, the likely significance of this with respect to interpreting the bulk properties of these materials has apparently been overlooked in the preceding three decades. Additionally, their regular spatial distribution and morphology, along with the lack of correlation between $\Gamma_{4}^{+}$ and $\Gamma_{1}^{+}$(1), confirms that they are not artefacts from chemical inhomogeneities but rather reflect an intrinsic and robust microstructure. 

From the observed strain profiles, only negative $\Gamma_{4}^{+}$ is present at 100 K which constrains the domain configurations. The absence of positive $\Gamma_{4}^{+}$ implies that the domain sequences must be of the form X$_{3}^{+}$($-a;a$)$|$X$_{3}^{+}$($0;a$)$|$X$_{3}^{+}$($a;a$) or X$_{3}^{+}$($-a;-a$)$|$X$_{3}^{+}$($0;-a$)$|$X$_{3}^{+}$($a;-a$). These LTO-like inclusions may locally disrupt the CDW-stabilising LTT tilt pattern, constraining the CDW correlation length and potentially explaining why it does not fully saturate in LESCO \cite{lee_generic_2022}, and suggesting that similar mesoscale heterogeneity could influence CDW correlations in other members of the cuprate family that exhibit comparable pseudosymmetry. Such fine structural heterogeneity likely actively shapes CDW order and may provide filamentary pathways for localised or lower-dimensional superconductivity \cite{gofryk_local_2014, xiao_evidence_2012, venditti_charge-density_2023}. Recent studies have emphasised the importance of structural domain walls, particularly those associated with LTO-like distortions, in governing the formation and stability of CDW order in cuprates more broadly \cite{chen_charge_2019}. However, a detailed description of the underlying microstructure has so far remained elusive. The present results help address this gap, suggesting that fine structural heterogeneity may provide a natural framework for the microscopic coexistence of CDW order, superconductivity, and even antiferromagnetic order \cite{pratt_coexistence_2009, choi_spatially_2020}.

Our interpretation follows the established association of CDW order occurring within the LTT phase and superconductivity within the LTO phase in 214 cuprates; however, direct spatial correlation between structure and electronic response remains an important open problem. We emphasise that the present work is focused on resolving the bulk microstructure. Given the correlation length associated with SC and CDW order parameters are in the region of 2 -- 15 nm in cuprates, we anticipate that the symmetry breaking associated with the domain wall structures shape the competition between SC and CDW, rather than being shaped by it.  Put another way, the SC or CDW order parameters will feel the domain wall symmetry as “bulk”.  However, as ferroelastic strains are decreased, the domain wall energy and width, generally having a quadratic dependence on this \cite{Salje2000}, will fall rapidly. Reductions in the LTO transition temperatures and hence ferroelastic strains are evident across the 214- cuprate family members, where they are strongly influenced by tolerance factor, variance and doping effects \cite{Herlihy2025}.

The 214s are far from the only cuprates expected to exhibit such domain wall structures.  Indeed, a symmetry analysis of all aristotypical parent structures of the main cuprate superconductors reveal a striking amount of generality, all being based on $4/mmm$ lattice symmetry. For the purposes of our current discussion, it is helpful to divide them into two classes. 

Class (A) containing 214s (e.g. LESCO, LSCO, LBCO), the Bi-12$n$[$n$+1] (Bi$_2$Sr$_2$CuO$_6$, BiSr$_2$CaCu$_2$O$_8$, Bi$_2$Sr$_2$Ca$_2$Cu$_3$O$_{10}$) and Tl-12$n$[$n$+1] (Tl$_2$Ba$_2$CuO$_6$, Tl$_2$Ba$_2$CaCu$_2$O$_8$, Tl$_2$Ba$_2$Ca$_2$Cu$_3$O$_{10}$), for which a parent structure with symmetry $I4/mmm$  can be found. Examples in the literature for all families exist in which structural distortions, corresponding to CuO$_2$ plane tilting (transforming as X$_3^+$ or X$_4^+$), are reported. These distortions couple to symmetry breaking strains along a direction lying in-between the O-Cu-O planes, (the $\Gamma_4^+$ strain channel).

Class (B) Hg12n[n+1] (HgBa$_2$CuO$_4$, HgBa$_2$CaCu$_2$O$_6$, HgBa$_2$Ca$_2$Cu$_3$O$_8$) and YBCO (YBa$_2$Cu$_3$O$_{7-x}$) with a $P4/mmm$ parent aristotype.  Here, where observed, symmetry breaking seems to be predominantly associated with anion ordering which couples to symmetry breaking strains lying along the Cu-O bonds ($\Gamma_2^+$ strain channel).

Class (A) contain the majority of high-Tc compounds.  Given the symmetry breaking strains transform as the same representation of the parent structure, and the parent structures will have both similar elastic constants and low-energy twin planes, we would predict our result to be general in this class, while noting the discussion above on the scaling of domain wall widths. The variation of calculated $\Gamma_4^+$  strains and therefore expected domain sizes appears to range from 7 $\times$ 10$^{-3}$ in LESCO to  3 $\times$ 10$^{-3}$ in Tl-2201 \cite{Tl2Ba2CuO6} all the way down to 3 $\times$ 10$^{-4}$ in   Bi-2223 \cite{Bi2Sr2Ca2Cu3O10}.   This class is also likely to be a relevant description to a broad range of electronic materials with layered perovskite structures, including superconductors Sr$_2$RuO$_4$ \cite{maeno_thirty_2024} and La$_3$Ni$_2$O$_7$ \cite{puphal_unconventional_2024, chen_polymorphism_2024, wang_long-range_2024} , which can exhibit complex structural phase transitions driven by octahedral rotations and tiltings.

For class (B), it may also be fruitful to investigate these using scanning 3DXRD but we would anticipate the most striking contrast would  be in $\Gamma_2^+$ spatially resolved strains.  We note also that strains in YBCO are typically large, and of the order 0.01, suggesting a larger field of view may be required. Indeed, twin domains can be viewed optically in YBCO where they are many microns in size, and procedures exist for detwinning them under uniaxial compression \cite{YBCO_detwin}. On the other hand, symmetry breaking in HgBa$_2$CuO$_4$ has been interpreted as nanoscale correlations of atomic displacements, \cite{Osborn2024}, with no reported macroscopic strains \cite{Hg1201}, suggesting the lengths scales of any structural heterogeneity are well below the intrinsic limit of the technique.
While LESCO provides an important model system, the broader significance of this work lies in the capabilities of the experimental approach. Scanning 3DXRD enables direct visualisation of bulk microstructural heterogeneity with a level of detail that has not previously been accessible across such length scales. More generally, any phase transition that may be considered as ferroelastic or co-elastic \cite{Salje_book}, can be imaged using our implementation of scanning 3DXRD, provided the length scales of domain structures are upwards of 70 nm. This includes martensitic phase transitions in which strain is the primary order parameter and typically of a very large magnitude,  but it also includes almost all structural and magnetic phase transitions where strain (symmetry breaking or symmetric) invariably has a coupling to the primary order parameter. The ability to resolve such structures in three dimensions within bulk samples opens up new opportunities to revisit these systems and to quantitatively link micro- and mesostructure with macroscopic properties.

Unfortunately, direct tomographic mapping of weak CDW reflections is beyond current 3DXRD capabilities, but future improvements in beamline technology and methods could enable sub-100 nm bulk mapping of structure and CDW order. This presents a clear pathway for establishing direct, spatially resolved structure-electronic correlations, which would significantly advance our understanding of competing orders in complex oxides. Beyond diffraction techniques, the observed complexity and variation in these microstructural features highlight the need for local probe techniques capable of resolving electronic behaviour on comparable length scales. While techniques such as conductive atomic force microscopy (cAFM)  shows promise for certain materials, in the case of LESCO we found that anion migration currently prevents reliable interpretation, leaving the development of robust local probe methodologies an open challenge. Spatially resolved spectroscopic techniques such as nano-ARPES may also provide clues on the extent of the electronic heterogeneity, however as with cAFM, these will also be inherently surface-sensitive techniques.  More broadly, the results underscore that establishing the intrinsic bulk structure is essential for interpreting electronic properties in systems with competing orders. This is particularly important in ferroelastic materials, where strain fields can relax at free surfaces, meaning that surface-sensitive probes may not directly reflect bulk behaviour. In this context, bulk-sensitive structural characterisation is not secondary but complementary, and necessary for a complete understanding of electronic measurements.

Ultimately, our direct visualisation of these microstructural heterogeneities within bulk samples prompts a re-evaluation of how material structure is conceptualised. Increasingly, such variations in the field of ferroic materials are seen not as simple interfaces but as potentially functional components \cite{salje_multiferroic_2010}. This raises a philosophical question: should we consider these extensive heterogeneities as distinct coexisting phases, or should we instead hold a more holistic view of material structure in which the intricate interplay of continuous and discrete structural variations governs macroscopic properties? New experimental techniques, such as the use of scanning three-dimensional X-ray diffraction that we have demonstrated here, will be instrumental in capturing the high-resolution structural data necessary to confront this challenge, enabling a more nuanced understanding of complex materials.

The data that support the findings of this article are openly available \cite{ESRF_data}, embargo periods may apply.

\begin{acknowledgments}
We thank Lauren Cane and Stephen Hayden for synthesis of the sample and Philipp Fahler-M\"unzer and Donald Evans for test cAFM measurements. E. L. thanks the University of Warwick for a PhD studentship through the Warwick Centre for Doctoral Training in Analytical Science. M. S. S. acknowledges the Royal Society for a fellowship (UF160265 $\&$ URF$\backslash$R$\backslash$231012) for funding. We acknowledge the European Synchrotron Radiation Facility (ESRF) for provision of synchrotron radiation facilities at beamline ID11 under proposal number HC-5590 \cite{ESRF_data}. 
\end{acknowledgments}

\bibliography{apssamp}

\begin{thebibliography}{74}%
\makeatletter
\providecommand \@ifxundefined [1]{%
 \@ifx{#1\undefined}
}%
\providecommand \@ifnum [1]{%
 \ifnum #1\expandafter \@firstoftwo
 \else \expandafter \@secondoftwo
 \fi
}%
\providecommand \@ifx [1]{%
 \ifx #1\expandafter \@firstoftwo
 \else \expandafter \@secondoftwo
 \fi
}%
\providecommand \natexlab [1]{#1}%
\providecommand \enquote  [1]{``#1''}%
\providecommand \bibnamefont  [1]{#1}%
\providecommand \bibfnamefont [1]{#1}%
\providecommand \citenamefont [1]{#1}%
\providecommand \href@noop [0]{\@secondoftwo}%
\providecommand \href [0]{\begingroup \@sanitize@url \@href}%
\providecommand \@href[1]{\@@startlink{#1}\@@href}%
\providecommand \@@href[1]{\endgroup#1\@@endlink}%
\providecommand \@sanitize@url [0]{\catcode `\\12\catcode `\$12\catcode
  `\&12\catcode `\#12\catcode `\^12\catcode `\_12\catcode `\%12\relax}%
\providecommand \@@startlink[1]{}%
\providecommand \@@endlink[0]{}%
\providecommand \url  [0]{\begingroup\@sanitize@url \@url }%
\providecommand \@url [1]{\endgroup\@href {#1}{\urlprefix }}%
\providecommand \urlprefix  [0]{URL }%
\providecommand \Eprint [0]{\href }%
\providecommand \doibase [0]{https://doi.org/}%
\providecommand \selectlanguage [0]{\@gobble}%
\providecommand \bibinfo  [0]{\@secondoftwo}%
\providecommand \bibfield  [0]{\@secondoftwo}%
\providecommand \translation [1]{[#1]}%
\providecommand \BibitemOpen [0]{}%
\providecommand \bibitemStop [0]{}%
\providecommand \bibitemNoStop [0]{.\EOS\space}%
\providecommand \EOS [0]{\spacefactor3000\relax}%
\providecommand \BibitemShut  [1]{\csname bibitem#1\endcsname}%
\let\auto@bib@innerbib\@empty
\bibitem [{\citenamefont {Seidel}\ \emph {et~al.}(2009)\citenamefont {Seidel},
  \citenamefont {Martin}, \citenamefont {He}, \citenamefont {Zhan},
  \citenamefont {Chu}, \citenamefont {Rother}, \citenamefont {Hawkridge},
  \citenamefont {Maksymovych}, \citenamefont {Yu}, \citenamefont {Gajek},
  \citenamefont {Balke}, \citenamefont {Kalinin}, \citenamefont {Gemming},
  \citenamefont {Wang}, \citenamefont {Catalan}, \citenamefont {Scott},
  \citenamefont {Spaldin}, \citenamefont {Orenstein},\ and\ \citenamefont
  {Ramesh}}]{seidel_conduction_2009}%
  \BibitemOpen
  \bibfield  {author} {\bibinfo {author} {\bibfnamefont {J.}~\bibnamefont
  {Seidel}}, \bibinfo {author} {\bibfnamefont {L.~W.}\ \bibnamefont {Martin}},
  \bibinfo {author} {\bibfnamefont {Q.}~\bibnamefont {He}}, \bibinfo {author}
  {\bibfnamefont {Q.}~\bibnamefont {Zhan}}, \bibinfo {author} {\bibfnamefont
  {Y.-H.}\ \bibnamefont {Chu}}, \bibinfo {author} {\bibfnamefont
  {A.}~\bibnamefont {Rother}}, \bibinfo {author} {\bibfnamefont {M.~E.}\
  \bibnamefont {Hawkridge}}, \bibinfo {author} {\bibfnamefont {P.}~\bibnamefont
  {Maksymovych}}, \bibinfo {author} {\bibfnamefont {P.}~\bibnamefont {Yu}},
  \bibinfo {author} {\bibfnamefont {M.}~\bibnamefont {Gajek}}, \bibinfo
  {author} {\bibfnamefont {N.}~\bibnamefont {Balke}}, \bibinfo {author}
  {\bibfnamefont {S.~V.}\ \bibnamefont {Kalinin}}, \bibinfo {author}
  {\bibfnamefont {S.}~\bibnamefont {Gemming}}, \bibinfo {author} {\bibfnamefont
  {F.}~\bibnamefont {Wang}}, \bibinfo {author} {\bibfnamefont {G.}~\bibnamefont
  {Catalan}}, \bibinfo {author} {\bibfnamefont {J.~F.}\ \bibnamefont {Scott}},
  \bibinfo {author} {\bibfnamefont {N.~A.}\ \bibnamefont {Spaldin}}, \bibinfo
  {author} {\bibfnamefont {J.}~\bibnamefont {Orenstein}},\ and\ \bibinfo
  {author} {\bibfnamefont {R.}~\bibnamefont {Ramesh}},\ }\bibfield  {title}
  {\bibinfo {title} {Conduction at domain walls in oxide multiferroics},\
  }\href@noop {} {\bibfield  {journal} {\bibinfo  {journal} {Nature Materials}\
  }\textbf {\bibinfo {volume} {8}},\ \bibinfo {pages} {229} (\bibinfo {year}
  {2009})}\BibitemShut {NoStop}%
\bibitem [{\citenamefont {Zubko}\ \emph {et~al.}(2007)\citenamefont {Zubko},
  \citenamefont {Catalan}, \citenamefont {Buckley}, \citenamefont {Welche},\
  and\ \citenamefont {Scott}}]{zubko_strain-gradient-induced_2007}%
  \BibitemOpen
  \bibfield  {author} {\bibinfo {author} {\bibfnamefont {P.}~\bibnamefont
  {Zubko}}, \bibinfo {author} {\bibfnamefont {G.}~\bibnamefont {Catalan}},
  \bibinfo {author} {\bibfnamefont {A.}~\bibnamefont {Buckley}}, \bibinfo
  {author} {\bibfnamefont {P.~R.~L.}\ \bibnamefont {Welche}},\ and\ \bibinfo
  {author} {\bibfnamefont {J.~F.}\ \bibnamefont {Scott}},\ }\bibfield  {title}
  {\bibinfo {title} {Strain-{Gradient}-{Induced} {Polarization} in
  {SrTiO}$_{\textrm{3}}$ {Single} {Crystals}},\ }\href@noop {} {\bibfield
  {journal} {\bibinfo  {journal} {Physical Review Letters}\ }\textbf {\bibinfo
  {volume} {99}},\ \bibinfo {pages} {167601} (\bibinfo {year}
  {2007})}\BibitemShut {NoStop}%
\bibitem [{\citenamefont {Tao}\ \emph {et~al.}(2024)\citenamefont {Tao},
  \citenamefont {Jiang}, \citenamefont {Chen}, \citenamefont {Zhang},
  \citenamefont {Cao}, \citenamefont {Yao}, \citenamefont {Chen}, \citenamefont
  {Ye},\ and\ \citenamefont {Ma}}]{tao_ferroelectric_2024}%
  \BibitemOpen
  \bibfield  {author} {\bibinfo {author} {\bibfnamefont {A.}~\bibnamefont
  {Tao}}, \bibinfo {author} {\bibfnamefont {Y.}~\bibnamefont {Jiang}}, \bibinfo
  {author} {\bibfnamefont {S.}~\bibnamefont {Chen}}, \bibinfo {author}
  {\bibfnamefont {Y.}~\bibnamefont {Zhang}}, \bibinfo {author} {\bibfnamefont
  {Y.}~\bibnamefont {Cao}}, \bibinfo {author} {\bibfnamefont {T.}~\bibnamefont
  {Yao}}, \bibinfo {author} {\bibfnamefont {C.}~\bibnamefont {Chen}}, \bibinfo
  {author} {\bibfnamefont {H.}~\bibnamefont {Ye}},\ and\ \bibinfo {author}
  {\bibfnamefont {X.-L.}\ \bibnamefont {Ma}},\ }\bibfield  {title} {\bibinfo
  {title} {Ferroelectric polarization and magnetic structure at domain walls in
  a multiferroic film},\ }\href@noop {} {\bibfield  {journal} {\bibinfo
  {journal} {Nature Communications}\ }\textbf {\bibinfo {volume} {15}},\
  \bibinfo {pages} {6099} (\bibinfo {year} {2024})}\BibitemShut {NoStop}%
\bibitem [{\citenamefont {Dagotto}(2005)}]{dagotto_complexity_2005}%
  \BibitemOpen
  \bibfield  {author} {\bibinfo {author} {\bibfnamefont {E.}~\bibnamefont
  {Dagotto}},\ }\bibfield  {title} {\bibinfo {title} {Complexity in {Strongly}
  {Correlated} {Electronic} {Systems}},\ }\href@noop {} {\bibfield  {journal}
  {\bibinfo  {journal} {Science}\ }\textbf {\bibinfo {volume} {309}},\ \bibinfo
  {pages} {257} (\bibinfo {year} {2005})}\BibitemShut {NoStop}%
\bibitem [{\citenamefont {Campi}\ \emph {et~al.}(2015)\citenamefont {Campi},
  \citenamefont {Bianconi}, \citenamefont {Poccia}, \citenamefont {Bianconi},
  \citenamefont {Barba}, \citenamefont {Arrighetti}, \citenamefont {Innocenti},
  \citenamefont {Karpinski}, \citenamefont {Zhigadlo}, \citenamefont {Kazakov},
  \citenamefont {Burghammer}, \citenamefont {Zimmermann}, \citenamefont
  {Sprung},\ and\ \citenamefont {Ricci}}]{campi_inhomogeneity_2015}%
  \BibitemOpen
  \bibfield  {author} {\bibinfo {author} {\bibfnamefont {G.}~\bibnamefont
  {Campi}}, \bibinfo {author} {\bibfnamefont {A.}~\bibnamefont {Bianconi}},
  \bibinfo {author} {\bibfnamefont {N.}~\bibnamefont {Poccia}}, \bibinfo
  {author} {\bibfnamefont {G.}~\bibnamefont {Bianconi}}, \bibinfo {author}
  {\bibfnamefont {L.}~\bibnamefont {Barba}}, \bibinfo {author} {\bibfnamefont
  {G.}~\bibnamefont {Arrighetti}}, \bibinfo {author} {\bibfnamefont
  {D.}~\bibnamefont {Innocenti}}, \bibinfo {author} {\bibfnamefont
  {J.}~\bibnamefont {Karpinski}}, \bibinfo {author} {\bibfnamefont {N.~D.}\
  \bibnamefont {Zhigadlo}}, \bibinfo {author} {\bibfnamefont {S.~M.}\
  \bibnamefont {Kazakov}}, \bibinfo {author} {\bibfnamefont {M.}~\bibnamefont
  {Burghammer}}, \bibinfo {author} {\bibfnamefont {M.~v.}\ \bibnamefont
  {Zimmermann}}, \bibinfo {author} {\bibfnamefont {M.}~\bibnamefont {Sprung}},\
  and\ \bibinfo {author} {\bibfnamefont {A.}~\bibnamefont {Ricci}},\ }\bibfield
   {title} {\bibinfo {title} {Inhomogeneity of charge-density-wave order and
  quenched disorder in a high-{$T_C$} superconductor},\ }\href@noop {}
  {\bibfield  {journal} {\bibinfo  {journal} {Nature}\ }\textbf {\bibinfo
  {volume} {525}},\ \bibinfo {pages} {359} (\bibinfo {year}
  {2015})}\BibitemShut {NoStop}%
\bibitem [{\citenamefont {Uehara}\ \emph {et~al.}(1999)\citenamefont {Uehara},
  \citenamefont {Mori}, \citenamefont {Chen},\ and\ \citenamefont
  {Cheong}}]{uehara_percolative_1999}%
  \BibitemOpen
  \bibfield  {author} {\bibinfo {author} {\bibfnamefont {M.}~\bibnamefont
  {Uehara}}, \bibinfo {author} {\bibfnamefont {S.}~\bibnamefont {Mori}},
  \bibinfo {author} {\bibfnamefont {C.~H.}\ \bibnamefont {Chen}},\ and\
  \bibinfo {author} {\bibfnamefont {S.-W.}\ \bibnamefont {Cheong}},\ }\bibfield
   {title} {\bibinfo {title} {Percolative phase separation underlies colossal
  magnetoresistance in mixed-valent manganites},\ }\href@noop {} {\bibfield
  {journal} {\bibinfo  {journal} {Nature}\ }\textbf {\bibinfo {volume} {399}},\
  \bibinfo {pages} {560} (\bibinfo {year} {1999})}\BibitemShut {NoStop}%
\bibitem [{\citenamefont {Tranquada}\ \emph {et~al.}(1995)\citenamefont
  {Tranquada}, \citenamefont {Sternlieb}, \citenamefont {Axe}, \citenamefont
  {Nakamura},\ and\ \citenamefont {Uchida}}]{tranquada_evidence_1995}%
  \BibitemOpen
  \bibfield  {author} {\bibinfo {author} {\bibfnamefont {J.~M.}\ \bibnamefont
  {Tranquada}}, \bibinfo {author} {\bibfnamefont {B.~J.}\ \bibnamefont
  {Sternlieb}}, \bibinfo {author} {\bibfnamefont {J.~D.}\ \bibnamefont {Axe}},
  \bibinfo {author} {\bibfnamefont {Y.}~\bibnamefont {Nakamura}},\ and\
  \bibinfo {author} {\bibfnamefont {S.}~\bibnamefont {Uchida}},\ }\bibfield
  {title} {\bibinfo {title} {Evidence for stripe correlations of spins and
  holes in copper oxide superconductors},\ }\href@noop {} {\bibfield  {journal}
  {\bibinfo  {journal} {Nature}\ }\textbf {\bibinfo {volume} {375}},\ \bibinfo
  {pages} {561} (\bibinfo {year} {1995})}\BibitemShut {NoStop}%
\bibitem [{\citenamefont {Hinkov}\ \emph {et~al.}(2008)\citenamefont {Hinkov},
  \citenamefont {Haug}, \citenamefont {Fauque}, \citenamefont {Bourges},
  \citenamefont {Sidis}, \citenamefont {Ivanov}, \citenamefont {Bernhard},
  \citenamefont {Lin},\ and\ \citenamefont {Keimer}}]{hinkov_electronic_2008}%
  \BibitemOpen
  \bibfield  {author} {\bibinfo {author} {\bibfnamefont {V.}~\bibnamefont
  {Hinkov}}, \bibinfo {author} {\bibfnamefont {D.}~\bibnamefont {Haug}},
  \bibinfo {author} {\bibfnamefont {B.}~\bibnamefont {Fauque}}, \bibinfo
  {author} {\bibfnamefont {P.}~\bibnamefont {Bourges}}, \bibinfo {author}
  {\bibfnamefont {Y.}~\bibnamefont {Sidis}}, \bibinfo {author} {\bibfnamefont
  {A.}~\bibnamefont {Ivanov}}, \bibinfo {author} {\bibfnamefont
  {C.}~\bibnamefont {Bernhard}}, \bibinfo {author} {\bibfnamefont {C.~T.}\
  \bibnamefont {Lin}},\ and\ \bibinfo {author} {\bibfnamefont {B.}~\bibnamefont
  {Keimer}},\ }\bibfield  {title} {\bibinfo {title} {Electronic {Liquid}
  {Crystal} {State} in the {High}-{Temperature} {Superconductor}
  yba$_2$cu$_3$o$_{6.45}$},\ }\href@noop {} {\bibfield  {journal} {\bibinfo
  {journal} {Science}\ }\textbf {\bibinfo {volume} {319}},\ \bibinfo {pages}
  {597} (\bibinfo {year} {2008})}\BibitemShut {NoStop}%
\bibitem [{\citenamefont {Phillips}(1997)}]{phillips_filamentary_1997}%
  \BibitemOpen
  \bibfield  {author} {\bibinfo {author} {\bibfnamefont {J.~C.}\ \bibnamefont
  {Phillips}},\ }\bibfield  {title} {\bibinfo {title} {Filamentary
  microstructure and linear temperature dependence of normal state transport in
  optimized high temperature-superconductors},\ }\href@noop {} {\bibfield
  {journal} {\bibinfo  {journal} {Proceedings of the National Academy of
  Sciences}\ }\textbf {\bibinfo {volume} {94}},\ \bibinfo {pages} {12771}
  (\bibinfo {year} {1997})}\BibitemShut {NoStop}%
\bibitem [{\citenamefont {Egami}\ and\ \citenamefont
  {Billinge}(1994)}]{egami_lattice_1994}%
  \BibitemOpen
  \bibfield  {author} {\bibinfo {author} {\bibfnamefont {T.}~\bibnamefont
  {Egami}}\ and\ \bibinfo {author} {\bibfnamefont {S.~J.~L.}\ \bibnamefont
  {Billinge}},\ }\bibfield  {title} {\bibinfo {title} {Lattice effects in high
  temperature superconductors},\ }\href
  {https://doi.org/10.1016/0079-6425(94)90005-1} {\bibfield  {journal}
  {\bibinfo  {journal} {Progress in Materials Science}\ }\textbf {\bibinfo
  {volume} {38}},\ \bibinfo {pages} {359} (\bibinfo {year} {1994})}\BibitemShut
  {NoStop}%
\bibitem [{\citenamefont {Saini}\ \emph {et~al.}(2001)\citenamefont {Saini},
  \citenamefont {Bianconi},\ and\ \citenamefont
  {Oyanagi}}]{saini_evidence_2001}%
  \BibitemOpen
  \bibfield  {author} {\bibinfo {author} {\bibfnamefont {N.~L.}\ \bibnamefont
  {Saini}}, \bibinfo {author} {\bibfnamefont {A.}~\bibnamefont {Bianconi}},\
  and\ \bibinfo {author} {\bibfnamefont {H.}~\bibnamefont {Oyanagi}},\
  }\bibfield  {title} {\bibinfo {title} {Evidence for {Critical} {Lattice}
  {Fluctuations} in the {High} \textit{{T}$_{\textrm{c}}$} {Cuprates}},\ }\href
  {https://doi.org/10.1143/JPSJ.70.2092} {\bibfield  {journal} {\bibinfo
  {journal} {Journal of the Physical Society of Japan}\ }\textbf {\bibinfo
  {volume} {70}},\ \bibinfo {pages} {2092} (\bibinfo {year}
  {2001})}\BibitemShut {NoStop}%
\bibitem [{\citenamefont {Pan}\ \emph {et~al.}(2001)\citenamefont {Pan},
  \citenamefont {O'Neal}, \citenamefont {Badzey}, \citenamefont {Chamon},
  \citenamefont {Ding}, \citenamefont {Engelbrecht}, \citenamefont {Wang},
  \citenamefont {Eisaki}, \citenamefont {Uchida}, \citenamefont {Gupta},
  \citenamefont {Ng}, \citenamefont {Hudson}, \citenamefont {Lang},\ and\
  \citenamefont {Davis}}]{pan_microscopic_2001}%
  \BibitemOpen
  \bibfield  {author} {\bibinfo {author} {\bibfnamefont {S.~H.}\ \bibnamefont
  {Pan}}, \bibinfo {author} {\bibfnamefont {J.~P.}\ \bibnamefont {O'Neal}},
  \bibinfo {author} {\bibfnamefont {R.~L.}\ \bibnamefont {Badzey}}, \bibinfo
  {author} {\bibfnamefont {C.}~\bibnamefont {Chamon}}, \bibinfo {author}
  {\bibfnamefont {H.}~\bibnamefont {Ding}}, \bibinfo {author} {\bibfnamefont
  {J.~R.}\ \bibnamefont {Engelbrecht}}, \bibinfo {author} {\bibfnamefont
  {Z.}~\bibnamefont {Wang}}, \bibinfo {author} {\bibfnamefont {H.}~\bibnamefont
  {Eisaki}}, \bibinfo {author} {\bibfnamefont {S.}~\bibnamefont {Uchida}},
  \bibinfo {author} {\bibfnamefont {A.~K.}\ \bibnamefont {Gupta}}, \bibinfo
  {author} {\bibfnamefont {K.-W.}\ \bibnamefont {Ng}}, \bibinfo {author}
  {\bibfnamefont {E.~W.}\ \bibnamefont {Hudson}}, \bibinfo {author}
  {\bibfnamefont {K.~M.}\ \bibnamefont {Lang}},\ and\ \bibinfo {author}
  {\bibfnamefont {J.~C.}\ \bibnamefont {Davis}},\ }\bibfield  {title} {\bibinfo
  {title} {Microscopic electronic inhomogeneity in the
  high-\textit{{T}$_{\textrm{c}}$} superconductor
  {Bi}$_{\textrm{2}}${Sr}$_{\textrm{2}}${CaCu}$_{\textrm{2}}${O}$_{\textrm{8+\textit{x}}}$},\
  }\href {https://doi.org/10.1038/35095012} {\bibfield  {journal} {\bibinfo
  {journal} {Nature}\ }\textbf {\bibinfo {volume} {413}},\ \bibinfo {pages}
  {282} (\bibinfo {year} {2001})}\BibitemShut {NoStop}%
\bibitem [{\citenamefont {Pasupathy}\ \emph {et~al.}(2008)\citenamefont
  {Pasupathy}, \citenamefont {Pushp}, \citenamefont {Gomes}, \citenamefont
  {Parker}, \citenamefont {Wen}, \citenamefont {Xu}, \citenamefont {Gu},
  \citenamefont {Ono}, \citenamefont {Ando},\ and\ \citenamefont
  {Yazdani}}]{pasupathy_electronic_2008}%
  \BibitemOpen
  \bibfield  {author} {\bibinfo {author} {\bibfnamefont {A.~N.}\ \bibnamefont
  {Pasupathy}}, \bibinfo {author} {\bibfnamefont {A.}~\bibnamefont {Pushp}},
  \bibinfo {author} {\bibfnamefont {K.~K.}\ \bibnamefont {Gomes}}, \bibinfo
  {author} {\bibfnamefont {C.~V.}\ \bibnamefont {Parker}}, \bibinfo {author}
  {\bibfnamefont {J.}~\bibnamefont {Wen}}, \bibinfo {author} {\bibfnamefont
  {Z.}~\bibnamefont {Xu}}, \bibinfo {author} {\bibfnamefont {G.}~\bibnamefont
  {Gu}}, \bibinfo {author} {\bibfnamefont {S.}~\bibnamefont {Ono}}, \bibinfo
  {author} {\bibfnamefont {Y.}~\bibnamefont {Ando}},\ and\ \bibinfo {author}
  {\bibfnamefont {A.}~\bibnamefont {Yazdani}},\ }\bibfield  {title} {\bibinfo
  {title} {Electronic {Origin} of the {Inhomogeneous} {Pairing} {Interaction}
  in the {High}-\textit{{T}$_{\textrm{c}}$} {Superconductor}
  {Bi}$_{\textrm{2}}${Sr}$_{\textrm{2}}${CaCu}$_{\textrm{2}}${O}$_{\textrm{8+\textit{d}}}$},\
  }\href {https://doi.org/10.1126/science.1154700} {\bibfield  {journal}
  {\bibinfo  {journal} {Science}\ }\textbf {\bibinfo {volume} {320}},\ \bibinfo
  {pages} {196} (\bibinfo {year} {2008})}\BibitemShut {NoStop}%
\bibitem [{\citenamefont {Moodenbaugh}\ \emph {et~al.}(1988)\citenamefont
  {Moodenbaugh}, \citenamefont {Xu}, \citenamefont {Suenaga}, \citenamefont
  {Folkerts},\ and\ \citenamefont
  {Shelton}}]{moodenbaugh_superconducting_1988}%
  \BibitemOpen
  \bibfield  {author} {\bibinfo {author} {\bibfnamefont {A.~R.}\ \bibnamefont
  {Moodenbaugh}}, \bibinfo {author} {\bibfnamefont {Y.}~\bibnamefont {Xu}},
  \bibinfo {author} {\bibfnamefont {M.}~\bibnamefont {Suenaga}}, \bibinfo
  {author} {\bibfnamefont {T.~J.}\ \bibnamefont {Folkerts}},\ and\ \bibinfo
  {author} {\bibfnamefont {R.~N.}\ \bibnamefont {Shelton}},\ }\bibfield
  {title} {\bibinfo {title} {Superconducting properties of
  {La}$_{\textrm{2-\textit{x}}}${Ba}$_{\textrm{\textit{x}}}${CuO}$_{\textrm{4}}$},\
  }\href@noop {} {\bibfield  {journal} {\bibinfo  {journal} {Physical Review
  B}\ }\textbf {\bibinfo {volume} {38}},\ \bibinfo {pages} {4596} (\bibinfo
  {year} {1988})}\BibitemShut {NoStop}%
\bibitem [{\citenamefont {Fink}\ \emph {et~al.}(2011)\citenamefont {Fink},
  \citenamefont {Soltwisch}, \citenamefont {Geck}, \citenamefont {Schierle},
  \citenamefont {Weschke},\ and\ \citenamefont
  {B\"{u}chner}}]{fink_phase_2011}%
  \BibitemOpen
  \bibfield  {author} {\bibinfo {author} {\bibfnamefont {J.}~\bibnamefont
  {Fink}}, \bibinfo {author} {\bibfnamefont {V.}~\bibnamefont {Soltwisch}},
  \bibinfo {author} {\bibfnamefont {J.}~\bibnamefont {Geck}}, \bibinfo {author}
  {\bibfnamefont {E.}~\bibnamefont {Schierle}}, \bibinfo {author}
  {\bibfnamefont {E.}~\bibnamefont {Weschke}},\ and\ \bibinfo {author}
  {\bibfnamefont {B.}~\bibnamefont {B\"{u}chner}},\ }\bibfield  {title}
  {\bibinfo {title} {Phase diagram of charge order in
  {La}$_{\textrm{1.8-\textit{x}}}${Eu}$_{\textrm{0.2}}${Sr}$_{\textrm{\textit{x}}}${CuO}$_{\textrm{4}}$
  from resonant soft x-ray diffraction},\ }\href@noop {} {\bibfield  {journal}
  {\bibinfo  {journal} {Physical Review B}\ }\textbf {\bibinfo {volume} {83}},\
  \bibinfo {pages} {092503} (\bibinfo {year} {2011})}\BibitemShut {NoStop}%
\bibitem [{\citenamefont {Tidey}\ \emph
  {et~al.}(2022{\natexlab{a}})\citenamefont {Tidey}, \citenamefont {Liu},
  \citenamefont {Lai}, \citenamefont {Chuang}, \citenamefont {Chen},
  \citenamefont {Cane}, \citenamefont {Lester}, \citenamefont {Petsch},
  \citenamefont {Herlihy}, \citenamefont {Simonov}, \citenamefont {Hayden},\
  and\ \citenamefont {Senn}}]{tidey_pronounced_2022}%
  \BibitemOpen
  \bibfield  {author} {\bibinfo {author} {\bibfnamefont {J.~P.}\ \bibnamefont
  {Tidey}}, \bibinfo {author} {\bibfnamefont {E.-P.}\ \bibnamefont {Liu}},
  \bibinfo {author} {\bibfnamefont {Y.-C.}\ \bibnamefont {Lai}}, \bibinfo
  {author} {\bibfnamefont {Y.-C.}\ \bibnamefont {Chuang}}, \bibinfo {author}
  {\bibfnamefont {W.-T.}\ \bibnamefont {Chen}}, \bibinfo {author}
  {\bibfnamefont {L.~J.}\ \bibnamefont {Cane}}, \bibinfo {author}
  {\bibfnamefont {C.}~\bibnamefont {Lester}}, \bibinfo {author} {\bibfnamefont
  {A.~N.~D.}\ \bibnamefont {Petsch}}, \bibinfo {author} {\bibfnamefont
  {A.}~\bibnamefont {Herlihy}}, \bibinfo {author} {\bibfnamefont
  {A.}~\bibnamefont {Simonov}}, \bibinfo {author} {\bibfnamefont {S.~M.}\
  \bibnamefont {Hayden}},\ and\ \bibinfo {author} {\bibfnamefont
  {M.}~\bibnamefont {Senn}},\ }\bibfield  {title} {\bibinfo {title} {Pronounced
  interplay between intrinsic phase-coexistence and octahedral tilt magnitude
  in hole-doped lanthanum cuprates},\ }\href@noop {} {\bibfield  {journal}
  {\bibinfo  {journal} {Scientific Reports}\ }\textbf {\bibinfo {volume}
  {12}},\ \bibinfo {pages} {14343} (\bibinfo {year}
  {2022}{\natexlab{a}})}\BibitemShut {NoStop}%
\bibitem [{\citenamefont {Tidey}\ \emph
  {et~al.}(2022{\natexlab{b}})\citenamefont {Tidey}, \citenamefont {Keegan},
  \citenamefont {Bristowe}, \citenamefont {Mostofi}, \citenamefont {Hong},
  \citenamefont {Chen}, \citenamefont {Chuang}, \citenamefont {Chen},\ and\
  \citenamefont {Senn}}]{tidey_structural_2022}%
  \BibitemOpen
  \bibfield  {author} {\bibinfo {author} {\bibfnamefont {J.~P.}\ \bibnamefont
  {Tidey}}, \bibinfo {author} {\bibfnamefont {C.}~\bibnamefont {Keegan}},
  \bibinfo {author} {\bibfnamefont {N.~C.}\ \bibnamefont {Bristowe}}, \bibinfo
  {author} {\bibfnamefont {A.~A.}\ \bibnamefont {Mostofi}}, \bibinfo {author}
  {\bibfnamefont {Z.-M.}\ \bibnamefont {Hong}}, \bibinfo {author}
  {\bibfnamefont {B.-H.}\ \bibnamefont {Chen}}, \bibinfo {author}
  {\bibfnamefont {Y.-C.}\ \bibnamefont {Chuang}}, \bibinfo {author}
  {\bibfnamefont {W.-T.}\ \bibnamefont {Chen}},\ and\ \bibinfo {author}
  {\bibfnamefont {M.~S.}\ \bibnamefont {Senn}},\ }\bibfield  {title} {\bibinfo
  {title} {Structural origins of the low-temperature orthorhombic to
  low-temperature tetragonal phase transition in high-{$T_C$} cuprates},\
  }\href@noop {} {\bibfield  {journal} {\bibinfo  {journal} {Physical Review
  B}\ }\textbf {\bibinfo {volume} {106}},\ \bibinfo {pages} {085112} (\bibinfo
  {year} {2022}{\natexlab{b}})}\BibitemShut {NoStop}%
\bibitem [{iso()}]{isotropy}%
  \BibitemOpen
  \href {https://iso.byu.edu/iso/isotropy.php} {}\bibinfo {note} {H. T. Stokes,
  D. M. Hatch, and B. J. Campbell, ISOTROPY Software Suite,
  \href{https://iso.byu.edu/iso/isotropy.php}{iso.byu.edu.}}\BibitemShut
  {Stop}%
\bibitem [{\citenamefont {{Campbell, B. J.}}\ \emph {et~al.}(2006)\citenamefont
  {{Campbell, B. J.}}, \citenamefont {{Stokes, H. T.}}, \citenamefont {{Tanner,
  D. E.}},\ and\ \citenamefont {{Hatch, D.
  M.}}}]{campbell_b_j_isodisplace_2006}%
  \BibitemOpen
  \bibfield  {author} {\bibinfo {author} {\bibnamefont {{Campbell, B. J.}}},
  \bibinfo {author} {\bibnamefont {{Stokes, H. T.}}}, \bibinfo {author}
  {\bibnamefont {{Tanner, D. E.}}},\ and\ \bibinfo {author} {\bibnamefont
  {{Hatch, D. M.}}},\ }\bibfield  {title} {\bibinfo {title} {{ISODISPLACE}: {A}
  web-based tool for exploring structural distortions.},\ }\href@noop {}
  {\bibfield  {journal} {\bibinfo  {journal} {Journal of Applied
  Crystallography}\ }\textbf {\bibinfo {volume} {39}},\ \bibinfo {pages} {607}
  (\bibinfo {year} {2006})}\BibitemShut {NoStop}%
\bibitem [{\citenamefont {Hanaguri}\ \emph {et~al.}(2004)\citenamefont
  {Hanaguri}, \citenamefont {Lupien}, \citenamefont {Kohsaka}, \citenamefont
  {Lee}, \citenamefont {Azuma}, \citenamefont {Takano}, \citenamefont
  {Takagi},\ and\ \citenamefont {Davis}}]{hanaguri_checkerboard_2004}%
  \BibitemOpen
  \bibfield  {author} {\bibinfo {author} {\bibfnamefont {T.}~\bibnamefont
  {Hanaguri}}, \bibinfo {author} {\bibfnamefont {C.}~\bibnamefont {Lupien}},
  \bibinfo {author} {\bibfnamefont {Y.}~\bibnamefont {Kohsaka}}, \bibinfo
  {author} {\bibfnamefont {D.-H.}\ \bibnamefont {Lee}}, \bibinfo {author}
  {\bibfnamefont {M.}~\bibnamefont {Azuma}}, \bibinfo {author} {\bibfnamefont
  {M.}~\bibnamefont {Takano}}, \bibinfo {author} {\bibfnamefont
  {H.}~\bibnamefont {Takagi}},\ and\ \bibinfo {author} {\bibfnamefont {J.~C.}\
  \bibnamefont {Davis}},\ }\bibfield  {title} {\bibinfo {title} {A checkerboard
  electronic crystal state in lightly hole-doped
  {Ca}$_{\textrm{2-\textit{x}}}${Na}$_{\textrm{\textit{x}}}${CuO}$_{\textrm{2}}${Cl}$_{\textrm{2}}$},\
  }\href@noop {} {\bibfield  {journal} {\bibinfo  {journal} {Nature}\ }\textbf
  {\bibinfo {volume} {430}},\ \bibinfo {pages} {1001} (\bibinfo {year}
  {2004})}\BibitemShut {NoStop}%
\bibitem [{\citenamefont {Berg}\ \emph {et~al.}(2007)\citenamefont {Berg},
  \citenamefont {Fradkin}, \citenamefont {Kim}, \citenamefont {Kivelson},
  \citenamefont {Oganesyan}, \citenamefont {Tranquada},\ and\ \citenamefont
  {Zhang}}]{berg_dynamical_2007}%
  \BibitemOpen
  \bibfield  {author} {\bibinfo {author} {\bibfnamefont {E.}~\bibnamefont
  {Berg}}, \bibinfo {author} {\bibfnamefont {E.}~\bibnamefont {Fradkin}},
  \bibinfo {author} {\bibfnamefont {E.-A.}\ \bibnamefont {Kim}}, \bibinfo
  {author} {\bibfnamefont {S.~A.}\ \bibnamefont {Kivelson}}, \bibinfo {author}
  {\bibfnamefont {V.}~\bibnamefont {Oganesyan}}, \bibinfo {author}
  {\bibfnamefont {J.~M.}\ \bibnamefont {Tranquada}},\ and\ \bibinfo {author}
  {\bibfnamefont {S.~C.}\ \bibnamefont {Zhang}},\ }\bibfield  {title} {\bibinfo
  {title} {Dynamical {Layer} {Decoupling} in a {Stripe}-{Ordered}
  {High}-{$T_C$} {Superconductor}},\ }\href@noop {} {\bibfield  {journal}
  {\bibinfo  {journal} {Physical Review Letters}\ }\textbf {\bibinfo {volume}
  {99}},\ \bibinfo {pages} {127003} (\bibinfo {year} {2007})}\BibitemShut
  {NoStop}%
\bibitem [{\citenamefont {Fujita}\ \emph {et~al.}(2004)\citenamefont {Fujita},
  \citenamefont {Goka}, \citenamefont {Yamada}, \citenamefont {Tranquada},\
  and\ \citenamefont {Regnault}}]{fujita_stripe_2004}%
  \BibitemOpen
  \bibfield  {author} {\bibinfo {author} {\bibfnamefont {M.}~\bibnamefont
  {Fujita}}, \bibinfo {author} {\bibfnamefont {H.}~\bibnamefont {Goka}},
  \bibinfo {author} {\bibfnamefont {K.}~\bibnamefont {Yamada}}, \bibinfo
  {author} {\bibfnamefont {J.~M.}\ \bibnamefont {Tranquada}},\ and\ \bibinfo
  {author} {\bibfnamefont {L.~P.}\ \bibnamefont {Regnault}},\ }\bibfield
  {title} {\bibinfo {title} {Stripe order, depinning, and fluctuations in
  {La}$_{\textrm{1.875}}${Ba}$_{\textrm{0.125}}${CuO}$_{\textrm{4}}$ and
  {La}$_{\textrm{1.875}}${Ba}$_{\textrm{0.075}}${Sr}$_{\textrm{0.050}}${CuO}$_{\textrm{4}}$},\
  }\href@noop {} {\bibfield  {journal} {\bibinfo  {journal} {Physical Review
  B}\ }\textbf {\bibinfo {volume} {70}},\ \bibinfo {pages} {104517} (\bibinfo
  {year} {2004})}\BibitemShut {NoStop}%
\bibitem [{\citenamefont {Sears}\ \emph {et~al.}(2023)\citenamefont {Sears},
  \citenamefont {Shen}, \citenamefont {Krogstad}, \citenamefont {Miao},
  \citenamefont {Bozin}, \citenamefont {Robinson}, \citenamefont {Gu},
  \citenamefont {Osborn}, \citenamefont {Rosenkranz}, \citenamefont
  {Tranquada},\ and\ \citenamefont {Dean}}]{sears_structure_2023}%
  \BibitemOpen
  \bibfield  {author} {\bibinfo {author} {\bibfnamefont {J.}~\bibnamefont
  {Sears}}, \bibinfo {author} {\bibfnamefont {Y.}~\bibnamefont {Shen}},
  \bibinfo {author} {\bibfnamefont {M.~J.}\ \bibnamefont {Krogstad}}, \bibinfo
  {author} {\bibfnamefont {H.}~\bibnamefont {Miao}}, \bibinfo {author}
  {\bibfnamefont {E.~S.}\ \bibnamefont {Bozin}}, \bibinfo {author}
  {\bibfnamefont {I.~K.}\ \bibnamefont {Robinson}}, \bibinfo {author}
  {\bibfnamefont {G.~D.}\ \bibnamefont {Gu}}, \bibinfo {author} {\bibfnamefont
  {R.}~\bibnamefont {Osborn}}, \bibinfo {author} {\bibfnamefont
  {S.}~\bibnamefont {Rosenkranz}}, \bibinfo {author} {\bibfnamefont {J.~M.}\
  \bibnamefont {Tranquada}},\ and\ \bibinfo {author} {\bibfnamefont {M.~P.~M.}\
  \bibnamefont {Dean}},\ }\bibfield  {title} {\bibinfo {title} {Structure of
  charge density waves in
  {La}$_{\textrm{1.875}}${Ba}$_{\textrm{0.125}}${CuO}$_{\textrm{4}}$},\
  }\href@noop {} {\bibfield  {journal} {\bibinfo  {journal} {Physical Review
  B}\ }\textbf {\bibinfo {volume} {107}},\ \bibinfo {pages} {115125} (\bibinfo
  {year} {2023})}\BibitemShut {NoStop}%
\bibitem [{\citenamefont {Poulsen}\ \emph {et~al.}(1997)\citenamefont
  {Poulsen}, \citenamefont {Garbe}, \citenamefont {Lorentzen}, \citenamefont
  {Juul~Jensen}, \citenamefont {Poulsen}, \citenamefont {Andersen},
  \citenamefont {Frello}, \citenamefont {Feidenhans'l},\ and\ \citenamefont
  {Graafsma}}]{poulsen_applications_1997}%
  \BibitemOpen
  \bibfield  {author} {\bibinfo {author} {\bibfnamefont {H.~F.}\ \bibnamefont
  {Poulsen}}, \bibinfo {author} {\bibfnamefont {S.}~\bibnamefont {Garbe}},
  \bibinfo {author} {\bibfnamefont {T.}~\bibnamefont {Lorentzen}}, \bibinfo
  {author} {\bibfnamefont {D.}~\bibnamefont {Juul~Jensen}}, \bibinfo {author}
  {\bibfnamefont {F.~W.}\ \bibnamefont {Poulsen}}, \bibinfo {author}
  {\bibfnamefont {N.~H.}\ \bibnamefont {Andersen}}, \bibinfo {author}
  {\bibfnamefont {T.}~\bibnamefont {Frello}}, \bibinfo {author} {\bibfnamefont
  {R.}~\bibnamefont {Feidenhans'l}},\ and\ \bibinfo {author} {\bibfnamefont
  {H.}~\bibnamefont {Graafsma}},\ }\bibfield  {title} {\bibinfo {title}
  {Applications of high-energy synchrotron radiation for structural studies of
  polycrystalline materials},\ }\href@noop {} {\bibfield  {journal} {\bibinfo
  {journal} {Journal of Synchrotron Radiation}\ }\textbf {\bibinfo {volume}
  {4}},\ \bibinfo {pages} {147} (\bibinfo {year} {1997})}\BibitemShut {NoStop}%
\bibitem [{\citenamefont {Wright}\ \emph {et~al.}(2020)\citenamefont {Wright},
  \citenamefont {Giacobbe},\ and\ \citenamefont {Majkut}}]{wright_new_2020}%
  \BibitemOpen
  \bibfield  {author} {\bibinfo {author} {\bibfnamefont {J.}~\bibnamefont
  {Wright}}, \bibinfo {author} {\bibfnamefont {C.}~\bibnamefont {Giacobbe}},\
  and\ \bibinfo {author} {\bibfnamefont {M.}~\bibnamefont {Majkut}},\
  }\bibfield  {title} {\bibinfo {title} {New opportunities at the materials
  science beamline at esrf to exploit high energy nano-focus {X}-ray beams},\
  }\href@noop {} {\bibfield  {journal} {\bibinfo  {journal} {Current Opinion in
  Solid State and Materials Science}\ }\textbf {\bibinfo {volume} {24}},\
  \bibinfo {pages} {100818} (\bibinfo {year} {2020})}\BibitemShut {NoStop}%
\bibitem [{ESR()}]{ESRF_data}%
  \BibitemOpen
  \href@noop {} {}\bibinfo {howpublished} {Chen, W. T., Ladbrook, E., Senn, M.
  S., and Simpson, S. (2027). Mapping the structural domains associated with
  the suppression of the 3D superconducting state in the high-Tc layered
  cuprate [Dataset]. European Synchrotron Radiation Facility.
  https://doi.org/10.15151/ESRF-ES-1734554406}\BibitemShut {NoStop}%
\bibitem [{\citenamefont {Henningsson}\ and\ \citenamefont
  {Hall}(2023)}]{henningsson_efficient_2023}%
  \BibitemOpen
  \bibfield  {author} {\bibinfo {author} {\bibfnamefont {A.}~\bibnamefont
  {Henningsson}}\ and\ \bibinfo {author} {\bibfnamefont {S.~A.}\ \bibnamefont
  {Hall}},\ }\bibfield  {title} {\bibinfo {title} {An efficient system matrix
  factorization method for scanning diffraction based strain tensor
  tomography},\ }\href@noop {} {\bibfield  {journal} {\bibinfo  {journal} {Acta
  Crystallographica. Section A, Foundations and Advances}\ }\textbf {\bibinfo
  {volume} {79}},\ \bibinfo {pages} {542} (\bibinfo {year} {2023})}\BibitemShut
  {NoStop}%
\bibitem [{Sup()}]{Supplemental_Material}%
  \BibitemOpen
  \href@noop {} {}\bibinfo {howpublished} {See Supplemental Material at [URL
  will be inserted by publisher] for details on 3DXRD measurements, data
  processing, symmetry-adapted strain analysis and measurement resolution. The
  Supplemental Material also contains Refs. \cite{tidey_pronounced_2022,
  wright_fable-3dxrdimaged11_2025, wright_using_2022, bonnin_impurity_2014,
  van_aarle_fast_2016, zhang_method_2019, kim_emergence_2023,
  henningsson_efficient_2023, smith_infrared_2019}}\BibitemShut {NoStop}%
\bibitem [{\citenamefont {Borb\"{e}ly}\ \emph {et~al.}(2014)\citenamefont
  {Borb\"{e}ly}, \citenamefont {Renversade},\ and\ \citenamefont
  {Kenesei}}]{borbely_calibration_2014}%
  \BibitemOpen
  \bibfield  {author} {\bibinfo {author} {\bibfnamefont {A.}~\bibnamefont
  {Borb\"{e}ly}}, \bibinfo {author} {\bibfnamefont {L.}~\bibnamefont
  {Renversade}},\ and\ \bibinfo {author} {\bibfnamefont {P.}~\bibnamefont
  {Kenesei}},\ }\bibfield  {title} {\bibinfo {title} {On the calibration of
  high-energy {X}-ray diffraction setups. {II}. {Assessing} the rotation axis
  and residual strains},\ }\href@noop {} {\bibfield  {journal} {\bibinfo
  {journal} {Journal of Applied Crystallography}\ }\textbf {\bibinfo {volume}
  {47}},\ \bibinfo {pages} {1585} (\bibinfo {year} {2014})}\BibitemShut
  {NoStop}%
\bibitem [{\citenamefont {Frison}\ \emph {et~al.}(2022)\citenamefont {Frison},
  \citenamefont {K\"{u}spert}, \citenamefont {Wang}, \citenamefont {Ivashko},
  \citenamefont {Zimmermann}, \citenamefont {Meven}, \citenamefont {Bucher},
  \citenamefont {Larsen}, \citenamefont {Niedermayer}, \citenamefont
  {Janoschek}, \citenamefont {Kurosawa}, \citenamefont {Momono}, \citenamefont
  {Oda}, \citenamefont {Christensen},\ and\ \citenamefont
  {Chang}}]{frison_crystal_2022}%
  \BibitemOpen
  \bibfield  {author} {\bibinfo {author} {\bibfnamefont {R.}~\bibnamefont
  {Frison}}, \bibinfo {author} {\bibfnamefont {J.}~\bibnamefont {K\"{u}spert}},
  \bibinfo {author} {\bibfnamefont {Q.}~\bibnamefont {Wang}}, \bibinfo {author}
  {\bibfnamefont {O.}~\bibnamefont {Ivashko}}, \bibinfo {author} {\bibfnamefont
  {M.~v.}\ \bibnamefont {Zimmermann}}, \bibinfo {author} {\bibfnamefont
  {M.}~\bibnamefont {Meven}}, \bibinfo {author} {\bibfnamefont
  {D.}~\bibnamefont {Bucher}}, \bibinfo {author} {\bibfnamefont
  {J.}~\bibnamefont {Larsen}}, \bibinfo {author} {\bibfnamefont
  {C.}~\bibnamefont {Niedermayer}}, \bibinfo {author} {\bibfnamefont
  {M.}~\bibnamefont {Janoschek}}, \bibinfo {author} {\bibfnamefont
  {T.}~\bibnamefont {Kurosawa}}, \bibinfo {author} {\bibfnamefont
  {N.}~\bibnamefont {Momono}}, \bibinfo {author} {\bibfnamefont
  {M.}~\bibnamefont {Oda}}, \bibinfo {author} {\bibfnamefont {N.~B.}\
  \bibnamefont {Christensen}},\ and\ \bibinfo {author} {\bibfnamefont
  {J.}~\bibnamefont {Chang}},\ }\bibfield  {title} {\bibinfo {title} {Crystal
  symmetry of stripe-ordered
  {La}$_{\textrm{1.88}}${Sr}$_{\textrm{0.12}}${CuO}$_{\textrm{4}}$},\ }\href
  {https://doi.org/10.1103/PhysRevB.105.224113} {\bibfield  {journal} {\bibinfo
   {journal} {Physical Review B}\ }\textbf {\bibinfo {volume} {105}},\ \bibinfo
  {pages} {224113} (\bibinfo {year} {2022})}\BibitemShut {NoStop}%
\bibitem [{\citenamefont {Hu}\ \emph {et~al.}(2025)\citenamefont {Hu},
  \citenamefont {Lozano}, \citenamefont {Ye}, \citenamefont {Li}, \citenamefont
  {Sears}, \citenamefont {Zaliznyak}, \citenamefont {Gu},\ and\ \citenamefont
  {Tranquada}}]{hu_charge_2025}%
  \BibitemOpen
  \bibfield  {author} {\bibinfo {author} {\bibfnamefont {X.}~\bibnamefont
  {Hu}}, \bibinfo {author} {\bibfnamefont {P.~M.}\ \bibnamefont {Lozano}},
  \bibinfo {author} {\bibfnamefont {F.}~\bibnamefont {Ye}}, \bibinfo {author}
  {\bibfnamefont {Q.}~\bibnamefont {Li}}, \bibinfo {author} {\bibfnamefont
  {J.}~\bibnamefont {Sears}}, \bibinfo {author} {\bibfnamefont {I.~A.}\
  \bibnamefont {Zaliznyak}}, \bibinfo {author} {\bibfnamefont {G.~D.}\
  \bibnamefont {Gu}},\ and\ \bibinfo {author} {\bibfnamefont {J.~M.}\
  \bibnamefont {Tranquada}},\ }\bibfield  {title} {\bibinfo {title} {Charge
  density waves and pinning by lattice anisotropy in 214 cuprates},\
  }\href@noop {} {\bibfield  {journal} {\bibinfo  {journal} {Phys. Rev. B}\
  }\textbf {\bibinfo {volume} {111}},\ \bibinfo {pages} {064504} (\bibinfo
  {year} {2025})}\BibitemShut {NoStop}%
\bibitem [{\citenamefont {Salje}\ and\ \citenamefont
  {Parlinski}(1991)}]{salje_microstructures_1991}%
  \BibitemOpen
  \bibfield  {author} {\bibinfo {author} {\bibfnamefont {E.}~\bibnamefont
  {Salje}}\ and\ \bibinfo {author} {\bibfnamefont {K.}~\bibnamefont
  {Parlinski}},\ }\bibfield  {title} {\bibinfo {title} {Microstructures in high
  t$_c$ superconductors},\ }\href@noop {} {\bibfield  {journal} {\bibinfo
  {journal} {Superconductor Science and Technology}\ }\textbf {\bibinfo
  {volume} {4}},\ \bibinfo {pages} {93} (\bibinfo {year} {1991})}\BibitemShut
  {NoStop}%
\bibitem [{\citenamefont {Croft}\ \emph {et~al.}(2014)\citenamefont {Croft},
  \citenamefont {Lester}, \citenamefont {Senn}, \citenamefont {Bombardi},\ and\
  \citenamefont {Hayden}}]{croft_charge_2014}%
  \BibitemOpen
  \bibfield  {author} {\bibinfo {author} {\bibfnamefont {T.~P.}\ \bibnamefont
  {Croft}}, \bibinfo {author} {\bibfnamefont {C.}~\bibnamefont {Lester}},
  \bibinfo {author} {\bibfnamefont {M.~S.}\ \bibnamefont {Senn}}, \bibinfo
  {author} {\bibfnamefont {A.}~\bibnamefont {Bombardi}},\ and\ \bibinfo
  {author} {\bibfnamefont {S.~M.}\ \bibnamefont {Hayden}},\ }\bibfield  {title}
  {\bibinfo {title} {Charge density wave fluctuations in
  la$_{2-x}$sr$_x$cuo$_4$ and their competition with superconductivity},\
  }\href@noop {} {\bibfield  {journal} {\bibinfo  {journal} {Phys. Rev. B}\
  }\textbf {\bibinfo {volume} {89}},\ \bibinfo {pages} {224513} (\bibinfo
  {year} {2014})}\BibitemShut {NoStop}%
\bibitem [{\citenamefont {Wang}\ \emph {et~al.}(2020)\citenamefont {Wang},
  \citenamefont {Horio}, \citenamefont {von Arx}, \citenamefont {Shen},
  \citenamefont {John~Mukkattukavil}, \citenamefont {Sassa}, \citenamefont
  {Ivashko}, \citenamefont {Matt}, \citenamefont {Pyon}, \citenamefont
  {Takayama}, \citenamefont {Takagi}, \citenamefont {Kurosawa}, \citenamefont
  {Momono}, \citenamefont {Oda}, \citenamefont {Adachi}, \citenamefont
  {Haidar}, \citenamefont {Koike}, \citenamefont {Tseng}, \citenamefont
  {Zhang}, \citenamefont {Zhao}, \citenamefont {Kummer}, \citenamefont
  {Garcia-Fernandez}, \citenamefont {Zhou}, \citenamefont {Christensen},
  \citenamefont {Ronnow}, \citenamefont {Schmitt},\ and\ \citenamefont
  {Chang}}]{wang_high-temperature_2020}%
  \BibitemOpen
  \bibfield  {author} {\bibinfo {author} {\bibfnamefont {Q.}~\bibnamefont
  {Wang}}, \bibinfo {author} {\bibfnamefont {M.}~\bibnamefont {Horio}},
  \bibinfo {author} {\bibfnamefont {K.}~\bibnamefont {von Arx}}, \bibinfo
  {author} {\bibfnamefont {Y.}~\bibnamefont {Shen}}, \bibinfo {author}
  {\bibfnamefont {D.}~\bibnamefont {John~Mukkattukavil}}, \bibinfo {author}
  {\bibfnamefont {Y.}~\bibnamefont {Sassa}}, \bibinfo {author} {\bibfnamefont
  {O.}~\bibnamefont {Ivashko}}, \bibinfo {author} {\bibfnamefont {C.-E.}\
  \bibnamefont {Matt}}, \bibinfo {author} {\bibfnamefont {S.}~\bibnamefont
  {Pyon}}, \bibinfo {author} {\bibfnamefont {T.}~\bibnamefont {Takayama}},
  \bibinfo {author} {\bibfnamefont {H.}~\bibnamefont {Takagi}}, \bibinfo
  {author} {\bibfnamefont {T.}~\bibnamefont {Kurosawa}}, \bibinfo {author}
  {\bibfnamefont {N.}~\bibnamefont {Momono}}, \bibinfo {author} {\bibfnamefont
  {M.}~\bibnamefont {Oda}}, \bibinfo {author} {\bibfnamefont {T.}~\bibnamefont
  {Adachi}}, \bibinfo {author} {\bibfnamefont {S.-M.}\ \bibnamefont {Haidar}},
  \bibinfo {author} {\bibfnamefont {Y.}~\bibnamefont {Koike}}, \bibinfo
  {author} {\bibfnamefont {Y.}~\bibnamefont {Tseng}}, \bibinfo {author}
  {\bibfnamefont {W.}~\bibnamefont {Zhang}}, \bibinfo {author} {\bibfnamefont
  {J.}~\bibnamefont {Zhao}}, \bibinfo {author} {\bibfnamefont {K.}~\bibnamefont
  {Kummer}}, \bibinfo {author} {\bibfnamefont {M.}~\bibnamefont
  {Garcia-Fernandez}}, \bibinfo {author} {\bibfnamefont {K.-J.}\ \bibnamefont
  {Zhou}}, \bibinfo {author} {\bibfnamefont {N.-B.}\ \bibnamefont
  {Christensen}}, \bibinfo {author} {\bibfnamefont {H.-M.}\ \bibnamefont
  {Ronnow}}, \bibinfo {author} {\bibfnamefont {T.}~\bibnamefont {Schmitt}},\
  and\ \bibinfo {author} {\bibfnamefont {J.}~\bibnamefont {Chang}},\ }\bibfield
   {title} {\bibinfo {title} {High-{Temperature} {Charge}-{Stripe}
  {Correlations} in
  {La}$_{\textrm{1.675}}${Eu}$_{\textrm{0.2}}${Sr}$_{\textrm{0.125}}${CuO}$_{\textrm{4}}$},\
  }\href@noop {} {\bibfield  {journal} {\bibinfo  {journal} {Physical Review
  Letters}\ }\textbf {\bibinfo {volume} {124}},\ \bibinfo {pages} {187002}
  (\bibinfo {year} {2020})}\BibitemShut {NoStop}%
\bibitem [{\citenamefont {Axe}\ \emph {et~al.}(1989)\citenamefont {Axe},
  \citenamefont {Moudden}, \citenamefont {Hohlwein}, \citenamefont {Cox},
  \citenamefont {Mohanty}, \citenamefont {Moodenbaugh},\ and\ \citenamefont
  {Xu}}]{axe_structural_1989}%
  \BibitemOpen
  \bibfield  {author} {\bibinfo {author} {\bibfnamefont {J.~D.}\ \bibnamefont
  {Axe}}, \bibinfo {author} {\bibfnamefont {A.~H.}\ \bibnamefont {Moudden}},
  \bibinfo {author} {\bibfnamefont {D.}~\bibnamefont {Hohlwein}}, \bibinfo
  {author} {\bibfnamefont {D.~E.}\ \bibnamefont {Cox}}, \bibinfo {author}
  {\bibfnamefont {K.~M.}\ \bibnamefont {Mohanty}}, \bibinfo {author}
  {\bibfnamefont {A.~R.}\ \bibnamefont {Moodenbaugh}},\ and\ \bibinfo {author}
  {\bibfnamefont {Y.}~\bibnamefont {Xu}},\ }\bibfield  {title} {\bibinfo
  {title} {Structural phase transformations and superconductivity in
  la$_{2-x}$ba$_x$cuo$_4$},\ }\href@noop {} {\bibfield  {journal} {\bibinfo
  {journal} {Physical Review Letters}\ }\textbf {\bibinfo {volume} {62}},\
  \bibinfo {pages} {2751} (\bibinfo {year} {1989})}\BibitemShut {NoStop}%
\bibitem [{\citenamefont {Crawford}\ \emph {et~al.}(1991)\citenamefont
  {Crawford}, \citenamefont {Harlow}, \citenamefont {McCarron}, \citenamefont
  {Farneth}, \citenamefont {Axe}, \citenamefont {Chou},\ and\ \citenamefont
  {Huang}}]{crawford_lattice_1991}%
  \BibitemOpen
  \bibfield  {author} {\bibinfo {author} {\bibfnamefont {M.~K.}\ \bibnamefont
  {Crawford}}, \bibinfo {author} {\bibfnamefont {R.~L.}\ \bibnamefont
  {Harlow}}, \bibinfo {author} {\bibfnamefont {E.~M.}\ \bibnamefont
  {McCarron}}, \bibinfo {author} {\bibfnamefont {W.~E.}\ \bibnamefont
  {Farneth}}, \bibinfo {author} {\bibfnamefont {J.~D.}\ \bibnamefont {Axe}},
  \bibinfo {author} {\bibfnamefont {H.}~\bibnamefont {Chou}},\ and\ \bibinfo
  {author} {\bibfnamefont {Q.}~\bibnamefont {Huang}},\ }\bibfield  {title}
  {\bibinfo {title} {Lattice instabilities and the effect of
  copper-oxygen-sheet distortions on superconductivity in doped
  la$_2$cuo$_4$},\ }\href@noop {} {\bibfield  {journal} {\bibinfo  {journal}
  {Physical Review B}\ }\textbf {\bibinfo {volume} {44}},\ \bibinfo {pages}
  {7749} (\bibinfo {year} {1991})}\BibitemShut {NoStop}%
\bibitem [{\citenamefont {Zhu}\ \emph {et~al.}(1994)\citenamefont {Zhu},
  \citenamefont {Moodenbaugh}, \citenamefont {Cai}, \citenamefont {Tafto},
  \citenamefont {Suenaga},\ and\ \citenamefont
  {Welch}}]{zhu_tetragonal-orthorhombic_1994}%
  \BibitemOpen
  \bibfield  {author} {\bibinfo {author} {\bibfnamefont {Y.}~\bibnamefont
  {Zhu}}, \bibinfo {author} {\bibfnamefont {A.~R.}\ \bibnamefont
  {Moodenbaugh}}, \bibinfo {author} {\bibfnamefont {Z.~X.}\ \bibnamefont
  {Cai}}, \bibinfo {author} {\bibfnamefont {J.}~\bibnamefont {Tafto}}, \bibinfo
  {author} {\bibfnamefont {M.}~\bibnamefont {Suenaga}},\ and\ \bibinfo {author}
  {\bibfnamefont {D.~O.}\ \bibnamefont {Welch}},\ }\bibfield  {title} {\bibinfo
  {title} {Tetragonal-{Orthorhombic} {Structural} {Modulation} at {Low}
  {Temperature} in
  {La}$_{\textrm{2-\textit{x}}}${Ba}$_{\textrm{\textit{x}}}${CuO}$_{\textrm{4}}$},\
  }\href@noop {} {\bibfield  {journal} {\bibinfo  {journal} {Physical Review
  Letters}\ }\textbf {\bibinfo {volume} {73}},\ \bibinfo {pages} {3026}
  (\bibinfo {year} {1994})}\BibitemShut {NoStop}%
\bibitem [{\citenamefont {Horibe}\ \emph {et~al.}(2000)\citenamefont {Horibe},
  \citenamefont {Inoue},\ and\ \citenamefont {Koyama}}]{horibe_direct_2000}%
  \BibitemOpen
  \bibfield  {author} {\bibinfo {author} {\bibfnamefont {Y.}~\bibnamefont
  {Horibe}}, \bibinfo {author} {\bibfnamefont {Y.}~\bibnamefont {Inoue}},\ and\
  \bibinfo {author} {\bibfnamefont {Y.}~\bibnamefont {Koyama}},\ }\bibfield
  {title} {\bibinfo {title} {Direct observation of dynamic local structure in
  {La}$_{\textrm{2-\textit{x}}}${Sr}$_{\textrm{\textit{x}}}${CuO}$_{\textrm{4}}$
  around $x$ = 0.12},\ }\href@noop {} {\bibfield  {journal} {\bibinfo
  {journal} {Physical Review B}\ }\textbf {\bibinfo {volume} {61}},\ \bibinfo
  {pages} {11922} (\bibinfo {year} {2000})}\BibitemShut {NoStop}%
\bibitem [{\citenamefont {Braden}\ \emph {et~al.}(1992)\citenamefont {Braden},
  \citenamefont {Heger}, \citenamefont {Schweiss}, \citenamefont {Fisk},
  \citenamefont {Gamayunov}, \citenamefont {Tanaka},\ and\ \citenamefont
  {Kojima}}]{braden_characterization_1992}%
  \BibitemOpen
  \bibfield  {author} {\bibinfo {author} {\bibfnamefont {M.}~\bibnamefont
  {Braden}}, \bibinfo {author} {\bibfnamefont {G.}~\bibnamefont {Heger}},
  \bibinfo {author} {\bibfnamefont {P.}~\bibnamefont {Schweiss}}, \bibinfo
  {author} {\bibfnamefont {Z.}~\bibnamefont {Fisk}}, \bibinfo {author}
  {\bibfnamefont {K.}~\bibnamefont {Gamayunov}}, \bibinfo {author}
  {\bibfnamefont {I.}~\bibnamefont {Tanaka}},\ and\ \bibinfo {author}
  {\bibfnamefont {H.}~\bibnamefont {Kojima}},\ }\bibfield  {title} {\bibinfo
  {title} {Characterization and structural analysis of twinned
  {La}$_{\textrm{2-\textit{x}}}${Sr}$_{\textrm{\textit{x}}}${CuO}$_{\textrm{4±\textit{delta}}}$
  crystals by neutron diffraction},\ }\href@noop {} {\bibfield  {journal}
  {\bibinfo  {journal} {Physica C: Superconductivity}\ }\textbf {\bibinfo
  {volume} {191}},\ \bibinfo {pages} {455} (\bibinfo {year}
  {1992})}\BibitemShut {NoStop}%
\bibitem [{\citenamefont {Horibe}\ \emph {et~al.}(1997)\citenamefont {Horibe},
  \citenamefont {Inoue},\ and\ \citenamefont
  {Koyama}}]{horibe_microstructure_1997}%
  \BibitemOpen
  \bibfield  {author} {\bibinfo {author} {\bibfnamefont {Y.}~\bibnamefont
  {Horibe}}, \bibinfo {author} {\bibfnamefont {Y.}~\bibnamefont {Inoue}},\ and\
  \bibinfo {author} {\bibfnamefont {K.}~\bibnamefont {Koyama}},\ }\bibfield
  {title} {\bibinfo {title} {Microstructure and {Tc}-suppression in
  {La}-cuprates},\ }\href@noop {} {\bibfield  {journal} {\bibinfo  {journal}
  {Physica C: Superconductivity}\ }\bibinfo {series} {Proceedings of the
  {International} {Conference} on {Materials} and {Mechanisms} of
  {Superconductivity} {High} {Temperature} {Superconductors} {V} {Part} {II}},\
  \textbf {\bibinfo {volume} {282-287}},\ \bibinfo {pages} {1071} (\bibinfo
  {year} {1997})}\BibitemShut {NoStop}%
\bibitem [{\citenamefont {Cai}\ and\ \citenamefont
  {Welch}(1994)}]{cai_theory_1994}%
  \BibitemOpen
  \bibfield  {author} {\bibinfo {author} {\bibfnamefont {Z.-X.}\ \bibnamefont
  {Cai}}\ and\ \bibinfo {author} {\bibfnamefont {D.~O.}\ \bibnamefont
  {Welch}},\ }\bibfield  {title} {\bibinfo {title} {Theory on the {Structure}
  of {Twin} {Boundaries} in
  {La}$_{\textrm{2-\textit{x}}}${Ba}$_{\textrm{\textit{x}}}${CuO}$_{\textrm{4}}$},\
  }\href@noop {} {\bibfield  {journal} {\bibinfo  {journal} {MRS Online
  Proceedings Library (OPL)}\ }\textbf {\bibinfo {volume} {357}},\ \bibinfo
  {pages} {453} (\bibinfo {year} {1994})}\BibitemShut {NoStop}%
\bibitem [{\citenamefont {Smith}\ \emph {et~al.}(2019)\citenamefont {Smith},
  \citenamefont {Nowadnick}, \citenamefont {Fan}, \citenamefont {Khatib},
  \citenamefont {Lim}, \citenamefont {Gao}, \citenamefont {Harms},
  \citenamefont {Neal}, \citenamefont {Kirkland}, \citenamefont {Martin},
  \citenamefont {Won}, \citenamefont {Raschke}, \citenamefont {Cheong},
  \citenamefont {Fennie}, \citenamefont {Carr}, \citenamefont {Bechtel},\ and\
  \citenamefont {Musfeldt}}]{smith_infrared_2019}%
  \BibitemOpen
  \bibfield  {author} {\bibinfo {author} {\bibfnamefont {K.~A.}\ \bibnamefont
  {Smith}}, \bibinfo {author} {\bibfnamefont {E.~A.}\ \bibnamefont
  {Nowadnick}}, \bibinfo {author} {\bibfnamefont {S.}~\bibnamefont {Fan}},
  \bibinfo {author} {\bibfnamefont {O.}~\bibnamefont {Khatib}}, \bibinfo
  {author} {\bibfnamefont {S.~J.}\ \bibnamefont {Lim}}, \bibinfo {author}
  {\bibfnamefont {B.}~\bibnamefont {Gao}}, \bibinfo {author} {\bibfnamefont
  {N.~C.}\ \bibnamefont {Harms}}, \bibinfo {author} {\bibfnamefont {S.~N.}\
  \bibnamefont {Neal}}, \bibinfo {author} {\bibfnamefont {J.~K.}\ \bibnamefont
  {Kirkland}}, \bibinfo {author} {\bibfnamefont {M.~C.}\ \bibnamefont
  {Martin}}, \bibinfo {author} {\bibfnamefont {C.~J.}\ \bibnamefont {Won}},
  \bibinfo {author} {\bibfnamefont {M.~B.}\ \bibnamefont {Raschke}}, \bibinfo
  {author} {\bibfnamefont {S.-W.}\ \bibnamefont {Cheong}}, \bibinfo {author}
  {\bibfnamefont {C.~J.}\ \bibnamefont {Fennie}}, \bibinfo {author}
  {\bibfnamefont {G.~L.}\ \bibnamefont {Carr}}, \bibinfo {author}
  {\bibfnamefont {H.~A.}\ \bibnamefont {Bechtel}},\ and\ \bibinfo {author}
  {\bibfnamefont {J.~L.}\ \bibnamefont {Musfeldt}},\ }\bibfield  {title}
  {\bibinfo {title} {Infrared nano-spectroscopy of ferroelastic domain walls in
  hybrid improper ferroelectric ca$_3$ti$_2$o$_7$},\ }\href@noop {} {\bibfield
  {journal} {\bibinfo  {journal} {Nature Communications}\ }\textbf {\bibinfo
  {volume} {10}},\ \bibinfo {pages} {5235} (\bibinfo {year}
  {2019})}\BibitemShut {NoStop}%
\bibitem [{\citenamefont {Yudin}\ \emph {et~al.}(2015)\citenamefont {Yudin},
  \citenamefont {Gureev}, \citenamefont {Sluka}, \citenamefont {Tagantsev},\
  and\ \citenamefont {Setter}}]{yudin_anomalously_2015}%
  \BibitemOpen
  \bibfield  {author} {\bibinfo {author} {\bibfnamefont {P.~V.}\ \bibnamefont
  {Yudin}}, \bibinfo {author} {\bibfnamefont {M.~Y.}\ \bibnamefont {Gureev}},
  \bibinfo {author} {\bibfnamefont {T.}~\bibnamefont {Sluka}}, \bibinfo
  {author} {\bibfnamefont {A.~K.}\ \bibnamefont {Tagantsev}},\ and\ \bibinfo
  {author} {\bibfnamefont {N.}~\bibnamefont {Setter}},\ }\bibfield  {title}
  {\bibinfo {title} {Anomalously thick domain walls in ferroelectrics},\
  }\href@noop {} {\bibfield  {journal} {\bibinfo  {journal} {Physical Review
  B}\ }\textbf {\bibinfo {volume} {91}},\ \bibinfo {pages} {060102} (\bibinfo
  {year} {2015})}\BibitemShut {NoStop}%
\bibitem [{\citenamefont {Lee}\ \emph {et~al.}(2017)\citenamefont {Lee},
  \citenamefont {Chang}, \citenamefont {Huang}, \citenamefont {Guo},
  \citenamefont {Gao}, \citenamefont {Chen}, \citenamefont {Cheong},\ and\
  \citenamefont {Chu}}]{lee_hidden_2017}%
  \BibitemOpen
  \bibfield  {author} {\bibinfo {author} {\bibfnamefont {M.~H.}\ \bibnamefont
  {Lee}}, \bibinfo {author} {\bibfnamefont {C.-P.}\ \bibnamefont {Chang}},
  \bibinfo {author} {\bibfnamefont {F.-T.}\ \bibnamefont {Huang}}, \bibinfo
  {author} {\bibfnamefont {G.~Y.}\ \bibnamefont {Guo}}, \bibinfo {author}
  {\bibfnamefont {B.}~\bibnamefont {Gao}}, \bibinfo {author} {\bibfnamefont
  {C.~H.}\ \bibnamefont {Chen}}, \bibinfo {author} {\bibfnamefont {S.-W.}\
  \bibnamefont {Cheong}},\ and\ \bibinfo {author} {\bibfnamefont {M.-W.}\
  \bibnamefont {Chu}},\ }\bibfield  {title} {\bibinfo {title} {Hidden antipolar
  order parameter and entangled neel-type charged domain walls in hybrid
  improper ferroelectrics},\ }\href@noop {} {\bibfield  {journal} {\bibinfo
  {journal} {Physical Review Letters}\ }\textbf {\bibinfo {volume} {119}},\
  \bibinfo {pages} {157601} (\bibinfo {year} {2017})}\BibitemShut {NoStop}%
\bibitem [{\citenamefont {Xia}\ \emph {et~al.}(2024)\citenamefont {Xia},
  \citenamefont {Bai}, \citenamefont {Chen}, \citenamefont {Yang},
  \citenamefont {Zhang}, \citenamefont {Yuan}, \citenamefont {Li},
  \citenamefont {Yang}, \citenamefont {Liu}, \citenamefont {Shi}, \citenamefont
  {Ma}, \citenamefont {Yang}, \citenamefont {He}, \citenamefont {Li},
  \citenamefont {Xi}, \citenamefont {Pi}, \citenamefont {Lv}, \citenamefont
  {Wang}, \citenamefont {Liu}, \citenamefont {Li}, \citenamefont {Zhou},
  \citenamefont {Liu}, \citenamefont {Chen}, \citenamefont {Shen},
  \citenamefont {Shen}, \citenamefont {Zhong}, \citenamefont {Wang},\ and\
  \citenamefont {Guo}}]{xia_giant_2024}%
  \BibitemOpen
  \bibfield  {author} {\bibinfo {author} {\bibfnamefont {W.}~\bibnamefont
  {Xia}}, \bibinfo {author} {\bibfnamefont {B.}~\bibnamefont {Bai}}, \bibinfo
  {author} {\bibfnamefont {X.}~\bibnamefont {Chen}}, \bibinfo {author}
  {\bibfnamefont {Y.}~\bibnamefont {Yang}}, \bibinfo {author} {\bibfnamefont
  {Y.}~\bibnamefont {Zhang}}, \bibinfo {author} {\bibfnamefont
  {J.}~\bibnamefont {Yuan}}, \bibinfo {author} {\bibfnamefont {Q.}~\bibnamefont
  {Li}}, \bibinfo {author} {\bibfnamefont {K.}~\bibnamefont {Yang}}, \bibinfo
  {author} {\bibfnamefont {X.}~\bibnamefont {Liu}}, \bibinfo {author}
  {\bibfnamefont {Y.}~\bibnamefont {Shi}}, \bibinfo {author} {\bibfnamefont
  {H.}~\bibnamefont {Ma}}, \bibinfo {author} {\bibfnamefont {H.}~\bibnamefont
  {Yang}}, \bibinfo {author} {\bibfnamefont {M.}~\bibnamefont {He}}, \bibinfo
  {author} {\bibfnamefont {L.}~\bibnamefont {Li}}, \bibinfo {author}
  {\bibfnamefont {C.}~\bibnamefont {Xi}}, \bibinfo {author} {\bibfnamefont
  {L.}~\bibnamefont {Pi}}, \bibinfo {author} {\bibfnamefont {X.}~\bibnamefont
  {Lv}}, \bibinfo {author} {\bibfnamefont {X.}~\bibnamefont {Wang}}, \bibinfo
  {author} {\bibfnamefont {X.}~\bibnamefont {Liu}}, \bibinfo {author}
  {\bibfnamefont {S.}~\bibnamefont {Li}}, \bibinfo {author} {\bibfnamefont
  {X.}~\bibnamefont {Zhou}}, \bibinfo {author} {\bibfnamefont {J.}~\bibnamefont
  {Liu}}, \bibinfo {author} {\bibfnamefont {Y.}~\bibnamefont {Chen}}, \bibinfo
  {author} {\bibfnamefont {J.}~\bibnamefont {Shen}}, \bibinfo {author}
  {\bibfnamefont {D.}~\bibnamefont {Shen}}, \bibinfo {author} {\bibfnamefont
  {Z.}~\bibnamefont {Zhong}}, \bibinfo {author} {\bibfnamefont
  {W.}~\bibnamefont {Wang}},\ and\ \bibinfo {author} {\bibfnamefont
  {Y.}~\bibnamefont {Guo}},\ }\bibfield  {title} {\bibinfo {title} {Giant
  {Domain} {Wall} {Anomalous} {Hall} {Effect} in a {Layered} {Antiferromagnet}
  {EuAl}$_{\textrm{2}}${Si}$_{\textrm{2}}$},\ }\href@noop {} {\bibfield
  {journal} {\bibinfo  {journal} {Physical Review Letters}\ }\textbf {\bibinfo
  {volume} {133}},\ \bibinfo {pages} {216602} (\bibinfo {year}
  {2024})}\BibitemShut {NoStop}%
\bibitem [{\citenamefont {Aird}\ and\ \citenamefont
  {Salje}(1998)}]{aird_sheet_1998}%
  \BibitemOpen
  \bibfield  {author} {\bibinfo {author} {\bibfnamefont {A.}~\bibnamefont
  {Aird}}\ and\ \bibinfo {author} {\bibfnamefont {E.~K.~H.}\ \bibnamefont
  {Salje}},\ }\bibfield  {title} {\bibinfo {title} {Sheet superconductivity in
  twin walls: experimental evidence of {W}{O}$_{\textrm{3-\textit{x}}}$},\
  }\href@noop {} {\bibfield  {journal} {\bibinfo  {journal} {Journal of
  Physics: Condensed Matter}\ }\textbf {\bibinfo {volume} {10}},\ \bibinfo
  {pages} {L377} (\bibinfo {year} {1998})}\BibitemShut {NoStop}%
\bibitem [{\citenamefont {Chen}\ \emph
  {et~al.}(1991{\natexlab{a}})\citenamefont {Chen}, \citenamefont {Cheong},
  \citenamefont {Werder}, \citenamefont {Cooper},\ and\ \citenamefont
  {Rupp}}]{chen_low_1991}%
  \BibitemOpen
  \bibfield  {author} {\bibinfo {author} {\bibfnamefont {C.~H.}\ \bibnamefont
  {Chen}}, \bibinfo {author} {\bibfnamefont {S.-W.}\ \bibnamefont {Cheong}},
  \bibinfo {author} {\bibfnamefont {D.~J.}\ \bibnamefont {Werder}}, \bibinfo
  {author} {\bibfnamefont {A.~S.}\ \bibnamefont {Cooper}},\ and\ \bibinfo
  {author} {\bibfnamefont {L.~W.}\ \bibnamefont {Rupp}},\ }\bibfield  {title}
  {\bibinfo {title} {Low temperature microstructure and phase transitions in
  la$_{2-x}$sr$_x$cuo$_4$ and la$_{2-x}$ba$_x$cuo$_4$},\ }\href@noop {}
  {\bibfield  {journal} {\bibinfo  {journal} {Physica C: Superconductivity}\
  }\textbf {\bibinfo {volume} {175}},\ \bibinfo {pages} {301} (\bibinfo {year}
  {1991}{\natexlab{a}})}\BibitemShut {NoStop}%
\bibitem [{\citenamefont {Chen}\ \emph
  {et~al.}(1991{\natexlab{b}})\citenamefont {Chen}, \citenamefont {Werder},
  \citenamefont {Cheong},\ and\ \citenamefont {Takagi}}]{chen_micro-twin_1991}%
  \BibitemOpen
  \bibfield  {author} {\bibinfo {author} {\bibfnamefont {C.~H.}\ \bibnamefont
  {Chen}}, \bibinfo {author} {\bibfnamefont {D.~J.}\ \bibnamefont {Werder}},
  \bibinfo {author} {\bibfnamefont {S.-W.}\ \bibnamefont {Cheong}},\ and\
  \bibinfo {author} {\bibfnamefont {H.}~\bibnamefont {Takagi}},\ }\bibfield
  {title} {\bibinfo {title} {Micro-twin and antiphase domain boundaries in the
  orthorhombic phase of la$_{2-x}$(sr,ba)$_x$cuo$_4$},\ }\href@noop {}
  {\bibfield  {journal} {\bibinfo  {journal} {Physica C: Superconductivity}\
  }\textbf {\bibinfo {volume} {183}},\ \bibinfo {pages} {121} (\bibinfo {year}
  {1991}{\natexlab{b}})}\BibitemShut {NoStop}%
\bibitem [{\citenamefont {Lee}\ \emph {et~al.}(2022)\citenamefont {Lee},
  \citenamefont {Huang}, \citenamefont {Johnson}, \citenamefont {Guo},
  \citenamefont {Husain}, \citenamefont {Mitrano}, \citenamefont {Lu},
  \citenamefont {Zakrzewski}, \citenamefont {de~la Pena}, \citenamefont {Peng},
  \citenamefont {Huang}, \citenamefont {Lee}, \citenamefont {Jang},
  \citenamefont {Lee}, \citenamefont {Joe}, \citenamefont {Doriese},
  \citenamefont {Szypryt}, \citenamefont {Swetz}, \citenamefont {Chi},
  \citenamefont {Aczel}, \citenamefont {MacDougall}, \citenamefont {Kivelson},
  \citenamefont {Fradkin},\ and\ \citenamefont {Abbamonte}}]{lee_generic_2022}%
  \BibitemOpen
  \bibfield  {author} {\bibinfo {author} {\bibfnamefont {S.}~\bibnamefont
  {Lee}}, \bibinfo {author} {\bibfnamefont {E.~W.}\ \bibnamefont {Huang}},
  \bibinfo {author} {\bibfnamefont {T.~A.}\ \bibnamefont {Johnson}}, \bibinfo
  {author} {\bibfnamefont {X.}~\bibnamefont {Guo}}, \bibinfo {author}
  {\bibfnamefont {A.~A.}\ \bibnamefont {Husain}}, \bibinfo {author}
  {\bibfnamefont {M.}~\bibnamefont {Mitrano}}, \bibinfo {author} {\bibfnamefont
  {K.}~\bibnamefont {Lu}}, \bibinfo {author} {\bibfnamefont {A.~V.}\
  \bibnamefont {Zakrzewski}}, \bibinfo {author} {\bibfnamefont {G.~A.}\
  \bibnamefont {de~la Pena}}, \bibinfo {author} {\bibfnamefont
  {Y.}~\bibnamefont {Peng}}, \bibinfo {author} {\bibfnamefont {H.}~\bibnamefont
  {Huang}}, \bibinfo {author} {\bibfnamefont {S.-J.}\ \bibnamefont {Lee}},
  \bibinfo {author} {\bibfnamefont {H.}~\bibnamefont {Jang}}, \bibinfo {author}
  {\bibfnamefont {J.-S.}\ \bibnamefont {Lee}}, \bibinfo {author} {\bibfnamefont
  {Y.~I.}\ \bibnamefont {Joe}}, \bibinfo {author} {\bibfnamefont {W.~B.}\
  \bibnamefont {Doriese}}, \bibinfo {author} {\bibfnamefont {P.}~\bibnamefont
  {Szypryt}}, \bibinfo {author} {\bibfnamefont {D.~S.}\ \bibnamefont {Swetz}},
  \bibinfo {author} {\bibfnamefont {S.}~\bibnamefont {Chi}}, \bibinfo {author}
  {\bibfnamefont {A.~A.}\ \bibnamefont {Aczel}}, \bibinfo {author}
  {\bibfnamefont {G.~J.}\ \bibnamefont {MacDougall}}, \bibinfo {author}
  {\bibfnamefont {S.~A.}\ \bibnamefont {Kivelson}}, \bibinfo {author}
  {\bibfnamefont {E.}~\bibnamefont {Fradkin}},\ and\ \bibinfo {author}
  {\bibfnamefont {P.}~\bibnamefont {Abbamonte}},\ }\bibfield  {title} {\bibinfo
  {title} {Generic character of charge and spin density waves in
  superconducting cuprates},\ }\href@noop {} {\bibfield  {journal} {\bibinfo
  {journal} {Proceedings of the National Academy of Sciences}\ }\textbf
  {\bibinfo {volume} {119}},\ \bibinfo {pages} {e2119429119} (\bibinfo {year}
  {2022})}\BibitemShut {NoStop}%
\bibitem [{\citenamefont {Gofryk}\ \emph {et~al.}(2014)\citenamefont {Gofryk},
  \citenamefont {Pan}, \citenamefont {Cantoni}, \citenamefont {Saparov},
  \citenamefont {Mitchell},\ and\ \citenamefont {Sefat}}]{gofryk_local_2014}%
  \BibitemOpen
  \bibfield  {author} {\bibinfo {author} {\bibfnamefont {K.}~\bibnamefont
  {Gofryk}}, \bibinfo {author} {\bibfnamefont {M.}~\bibnamefont {Pan}},
  \bibinfo {author} {\bibfnamefont {C.}~\bibnamefont {Cantoni}}, \bibinfo
  {author} {\bibfnamefont {B.}~\bibnamefont {Saparov}}, \bibinfo {author}
  {\bibfnamefont {J.~E.}\ \bibnamefont {Mitchell}},\ and\ \bibinfo {author}
  {\bibfnamefont {A.~S.}\ \bibnamefont {Sefat}},\ }\bibfield  {title} {\bibinfo
  {title} {Local {Inhomogeneity} and {Filamentary} {Superconductivity} in
  {Pr}-{Doped} {CaFe}$_{\textrm{2}}${As}$_{\textrm{2}}$},\ }\href@noop {}
  {\bibfield  {journal} {\bibinfo  {journal} {Physical Review Letters}\
  }\textbf {\bibinfo {volume} {112}},\ \bibinfo {pages} {047005} (\bibinfo
  {year} {2014})}\BibitemShut {NoStop}%
\bibitem [{\citenamefont {Xiao}\ \emph {et~al.}(2012)\citenamefont {Xiao},
  \citenamefont {Hu}, \citenamefont {Dioguardi}, \citenamefont {apRoberts
  Warren}, \citenamefont {Shockley}, \citenamefont {Crocker}, \citenamefont
  {Nisson}, \citenamefont {Viskadourakis}, \citenamefont {Tee}, \citenamefont
  {Radulov}, \citenamefont {Almasan}, \citenamefont {Curro},\ and\
  \citenamefont {Panagopoulos}}]{xiao_evidence_2012}%
  \BibitemOpen
  \bibfield  {author} {\bibinfo {author} {\bibfnamefont {H.}~\bibnamefont
  {Xiao}}, \bibinfo {author} {\bibfnamefont {T.}~\bibnamefont {Hu}}, \bibinfo
  {author} {\bibfnamefont {A.~P.}\ \bibnamefont {Dioguardi}}, \bibinfo {author}
  {\bibfnamefont {N.}~\bibnamefont {apRoberts Warren}}, \bibinfo {author}
  {\bibfnamefont {A.~C.}\ \bibnamefont {Shockley}}, \bibinfo {author}
  {\bibfnamefont {J.}~\bibnamefont {Crocker}}, \bibinfo {author} {\bibfnamefont
  {D.~M.}\ \bibnamefont {Nisson}}, \bibinfo {author} {\bibfnamefont
  {Z.}~\bibnamefont {Viskadourakis}}, \bibinfo {author} {\bibfnamefont
  {X.}~\bibnamefont {Tee}}, \bibinfo {author} {\bibfnamefont {I.}~\bibnamefont
  {Radulov}}, \bibinfo {author} {\bibfnamefont {C.~C.}\ \bibnamefont
  {Almasan}}, \bibinfo {author} {\bibfnamefont {N.~J.}\ \bibnamefont {Curro}},\
  and\ \bibinfo {author} {\bibfnamefont {C.}~\bibnamefont {Panagopoulos}},\
  }\bibfield  {title} {\bibinfo {title} {Evidence for filamentary
  superconductivity nucleated at antiphase domain walls in antiferromagnetic
  cafe$_2$as$_2$},\ }\href@noop {} {\bibfield  {journal} {\bibinfo  {journal}
  {Physical Review B}\ }\textbf {\bibinfo {volume} {85}},\ \bibinfo {pages}
  {024530} (\bibinfo {year} {2012})}\BibitemShut {NoStop}%
\bibitem [{\citenamefont {Venditti}\ and\ \citenamefont
  {Caprara}(2023)}]{venditti_charge-density_2023}%
  \BibitemOpen
  \bibfield  {author} {\bibinfo {author} {\bibfnamefont {G.}~\bibnamefont
  {Venditti}}\ and\ \bibinfo {author} {\bibfnamefont {S.}~\bibnamefont
  {Caprara}},\ }\bibfield  {title} {\bibinfo {title} {Charge-density waves vs.
  superconductivity: Some results and future perspectives},\ }\href@noop {}
  {\bibfield  {journal} {\bibinfo  {journal} {Condensed Matter}\ }\textbf
  {\bibinfo {volume} {8}},\ \bibinfo {pages} {54} (\bibinfo {year}
  {2023})}\BibitemShut {NoStop}%
\bibitem [{\citenamefont {Chen}\ \emph {et~al.}(2019)\citenamefont {Chen},
  \citenamefont {Mazzoli}, \citenamefont {Cao}, \citenamefont {Thampy},
  \citenamefont {Barbour}, \citenamefont {Hu}, \citenamefont {Lu},
  \citenamefont {Assefa}, \citenamefont {Miao}, \citenamefont {Fabbris},
  \citenamefont {Gu}, \citenamefont {Tranquada}, \citenamefont {Dean},
  \citenamefont {Wilkins},\ and\ \citenamefont {Robinson}}]{chen_charge_2019}%
  \BibitemOpen
  \bibfield  {author} {\bibinfo {author} {\bibfnamefont {X.~M.}\ \bibnamefont
  {Chen}}, \bibinfo {author} {\bibfnamefont {C.}~\bibnamefont {Mazzoli}},
  \bibinfo {author} {\bibfnamefont {Y.}~\bibnamefont {Cao}}, \bibinfo {author}
  {\bibfnamefont {V.}~\bibnamefont {Thampy}}, \bibinfo {author} {\bibfnamefont
  {A.~M.}\ \bibnamefont {Barbour}}, \bibinfo {author} {\bibfnamefont
  {W.}~\bibnamefont {Hu}}, \bibinfo {author} {\bibfnamefont {M.}~\bibnamefont
  {Lu}}, \bibinfo {author} {\bibfnamefont {T.~A.}\ \bibnamefont {Assefa}},
  \bibinfo {author} {\bibfnamefont {H.}~\bibnamefont {Miao}}, \bibinfo {author}
  {\bibfnamefont {G.}~\bibnamefont {Fabbris}}, \bibinfo {author} {\bibfnamefont
  {G.~D.}\ \bibnamefont {Gu}}, \bibinfo {author} {\bibfnamefont {J.~M.}\
  \bibnamefont {Tranquada}}, \bibinfo {author} {\bibfnamefont {M.~P.~M.}\
  \bibnamefont {Dean}}, \bibinfo {author} {\bibfnamefont {S.~B.}\ \bibnamefont
  {Wilkins}},\ and\ \bibinfo {author} {\bibfnamefont {I.~K.}\ \bibnamefont
  {Robinson}},\ }\bibfield  {title} {\bibinfo {title} {Charge density wave
  memory in a cuprate superconductor},\ }\href@noop {} {\bibfield  {journal}
  {\bibinfo  {journal} {Nature Communications}\ }\textbf {\bibinfo {volume}
  {10}},\ \bibinfo {pages} {1435} (\bibinfo {year} {2019})}\BibitemShut
  {NoStop}%
\bibitem [{\citenamefont {Pratt}\ \emph {et~al.}(2009)\citenamefont {Pratt},
  \citenamefont {Tian}, \citenamefont {Kreyssig}, \citenamefont {Zarestky},
  \citenamefont {Nandi}, \citenamefont {Ni}, \citenamefont {Bud'ko},
  \citenamefont {Canfield}, \citenamefont {Goldman},\ and\ \citenamefont
  {McQueeney}}]{pratt_coexistence_2009}%
  \BibitemOpen
  \bibfield  {author} {\bibinfo {author} {\bibfnamefont {D.~K.}\ \bibnamefont
  {Pratt}}, \bibinfo {author} {\bibfnamefont {W.}~\bibnamefont {Tian}},
  \bibinfo {author} {\bibfnamefont {A.}~\bibnamefont {Kreyssig}}, \bibinfo
  {author} {\bibfnamefont {J.~L.}\ \bibnamefont {Zarestky}}, \bibinfo {author}
  {\bibfnamefont {S.}~\bibnamefont {Nandi}}, \bibinfo {author} {\bibfnamefont
  {N.}~\bibnamefont {Ni}}, \bibinfo {author} {\bibfnamefont {S.~L.}\
  \bibnamefont {Bud'ko}}, \bibinfo {author} {\bibfnamefont {P.~C.}\
  \bibnamefont {Canfield}}, \bibinfo {author} {\bibfnamefont {A.~I.}\
  \bibnamefont {Goldman}},\ and\ \bibinfo {author} {\bibfnamefont {R.~J.}\
  \bibnamefont {McQueeney}},\ }\bibfield  {title} {\bibinfo {title}
  {Coexistence of competing antiferromagnetic and superconducting phases in the
  underdoped ba(fe$_{0.953}$co$_{0.047}$)$_2$as$_2$ compound using x-ray and
  neutron scattering techniques},\ }\href@noop {} {\bibfield  {journal}
  {\bibinfo  {journal} {Phys. Rev. Lett.}\ }\textbf {\bibinfo {volume} {103}},\
  \bibinfo {pages} {087001} (\bibinfo {year} {2009})}\BibitemShut {NoStop}%
\bibitem [{\citenamefont {Choi}\ \emph {et~al.}(2020)\citenamefont {Choi},
  \citenamefont {Ivashko}, \citenamefont {Blackburn}, \citenamefont {Liang},
  \citenamefont {Bonn}, \citenamefont {Hardy}, \citenamefont {Holmes},
  \citenamefont {Christensen}, \citenamefont {H\"{u}cker}, \citenamefont
  {Gerber}, \citenamefont {Gutowski}, \citenamefont {R\"{u}tt}, \citenamefont
  {Zimmermann}, \citenamefont {Forgan}, \citenamefont {Hayden},\ and\
  \citenamefont {Chang}}]{choi_spatially_2020}%
  \BibitemOpen
  \bibfield  {author} {\bibinfo {author} {\bibfnamefont {J.}~\bibnamefont
  {Choi}}, \bibinfo {author} {\bibfnamefont {O.}~\bibnamefont {Ivashko}},
  \bibinfo {author} {\bibfnamefont {E.}~\bibnamefont {Blackburn}}, \bibinfo
  {author} {\bibfnamefont {R.}~\bibnamefont {Liang}}, \bibinfo {author}
  {\bibfnamefont {D.~A.}\ \bibnamefont {Bonn}}, \bibinfo {author}
  {\bibfnamefont {W.~N.}\ \bibnamefont {Hardy}}, \bibinfo {author}
  {\bibfnamefont {A.~T.}\ \bibnamefont {Holmes}}, \bibinfo {author}
  {\bibfnamefont {N.~B.}\ \bibnamefont {Christensen}}, \bibinfo {author}
  {\bibfnamefont {M.}~\bibnamefont {H\"{u}cker}}, \bibinfo {author}
  {\bibfnamefont {S.}~\bibnamefont {Gerber}}, \bibinfo {author} {\bibfnamefont
  {O.}~\bibnamefont {Gutowski}}, \bibinfo {author} {\bibfnamefont
  {U.}~\bibnamefont {R\"{u}tt}}, \bibinfo {author} {\bibfnamefont {M.~v.}\
  \bibnamefont {Zimmermann}}, \bibinfo {author} {\bibfnamefont {E.~M.}\
  \bibnamefont {Forgan}}, \bibinfo {author} {\bibfnamefont {S.~M.}\
  \bibnamefont {Hayden}},\ and\ \bibinfo {author} {\bibfnamefont
  {J.}~\bibnamefont {Chang}},\ }\bibfield  {title} {\bibinfo {title} {Spatially
  inhomogeneous competition between superconductivity and the charge density
  wave in yba$_2$cu$_3$o$_{6.67}$},\ }\href@noop {} {\bibfield  {journal}
  {\bibinfo  {journal} {Nature Communications}\ }\textbf {\bibinfo {volume}
  {11}},\ \bibinfo {pages} {990} (\bibinfo {year} {2020})}\BibitemShut
  {NoStop}%
\bibitem [{\citenamefont {Salje}(2000)}]{Salje2000}%
  \BibitemOpen
  \bibfield  {author} {\bibinfo {author} {\bibfnamefont {E.~K.~H.}\
  \bibnamefont {Salje}},\ }\bibfield  {title} {\bibinfo {title}
  {Ferroelasticity},\ }\href {https://doi.org/10.1080/001075100181196}
  {\bibfield  {journal} {\bibinfo  {journal} {Contemporary Physics}\ }\textbf
  {\bibinfo {volume} {41}},\ \bibinfo {pages} {79} (\bibinfo {year}
  {2000})}\BibitemShut {NoStop}%
\bibitem [{\citenamefont {Herlihy}\ \emph {et~al.}(2025)\citenamefont
  {Herlihy}, \citenamefont {Chen}, \citenamefont {Ritter}, \citenamefont
  {Chuang},\ and\ \citenamefont {Senn}}]{Herlihy2025}%
  \BibitemOpen
  \bibfield  {author} {\bibinfo {author} {\bibfnamefont {A.}~\bibnamefont
  {Herlihy}}, \bibinfo {author} {\bibfnamefont {W.-T.}\ \bibnamefont {Chen}},
  \bibinfo {author} {\bibfnamefont {C.}~\bibnamefont {Ritter}}, \bibinfo
  {author} {\bibfnamefont {Y.-C.}\ \bibnamefont {Chuang}},\ and\ \bibinfo
  {author} {\bibfnamefont {M.~S.}\ \bibnamefont {Senn}},\ }\bibfield  {title}
  {\bibinfo {title} {Interplay between jahn–teller distortions and structural
  phase transitions in ruddlesden–poppers},\ }\href
  {https://doi.org/10.1021/jacs.5c00459} {\bibfield  {journal} {\bibinfo
  {journal} {Journal of the American Chemical Society}\ }\textbf {\bibinfo
  {volume} {147}},\ \bibinfo {pages} {7209} (\bibinfo {year}
  {2025})}\BibitemShut {NoStop}%
\bibitem [{\citenamefont {Parise}\ \emph {et~al.}(1988)\citenamefont {Parise},
  \citenamefont {Gopalakrishnan}, \citenamefont {Subramanian},\ and\
  \citenamefont {Sleight}}]{Tl2Ba2CuO6}%
  \BibitemOpen
  \bibfield  {author} {\bibinfo {author} {\bibfnamefont {J.}~\bibnamefont
  {Parise}}, \bibinfo {author} {\bibfnamefont {J.}~\bibnamefont
  {Gopalakrishnan}}, \bibinfo {author} {\bibfnamefont {M.}~\bibnamefont
  {Subramanian}},\ and\ \bibinfo {author} {\bibfnamefont {A.}~\bibnamefont
  {Sleight}},\ }\bibfield  {title} {\bibinfo {title} {Superconducting
  tl2ba2cuo6: The orthorhombic form},\ }\href
  {https://doi.org/https://doi.org/10.1016/0022-4596(88)90240-X} {\bibfield
  {journal} {\bibinfo  {journal} {Journal of Solid State Chemistry}\ }\textbf
  {\bibinfo {volume} {76}},\ \bibinfo {pages} {432} (\bibinfo {year}
  {1988})}\BibitemShut {NoStop}%
\bibitem [{\citenamefont {Shamray}\ \emph {et~al.}(2009)\citenamefont
  {Shamray}, \citenamefont {Mikhailova},\ and\ \citenamefont
  {Mitin}}]{Bi2Sr2Ca2Cu3O10}%
  \BibitemOpen
  \bibfield  {author} {\bibinfo {author} {\bibfnamefont {V.~F.}\ \bibnamefont
  {Shamray}}, \bibinfo {author} {\bibfnamefont {A.~B.}\ \bibnamefont
  {Mikhailova}},\ and\ \bibinfo {author} {\bibfnamefont {A.~V.}\ \bibnamefont
  {Mitin}},\ }\bibfield  {title} {\bibinfo {title} {Crystal structure and
  superconductivity of bi-2223},\ }\href
  {https://doi.org/10.1134/S1063774509040075} {\bibfield  {journal} {\bibinfo
  {journal} {Crystallography Reports}\ }\textbf {\bibinfo {volume} {54}},\
  \bibinfo {pages} {584} (\bibinfo {year} {2009})}\BibitemShut {NoStop}%
\bibitem [{\citenamefont {Maeno}\ \emph {et~al.}(2024)\citenamefont {Maeno},
  \citenamefont {Ikeda},\ and\ \citenamefont {Mattoni}}]{maeno_thirty_2024}%
  \BibitemOpen
  \bibfield  {author} {\bibinfo {author} {\bibfnamefont {Y.}~\bibnamefont
  {Maeno}}, \bibinfo {author} {\bibfnamefont {A.}~\bibnamefont {Ikeda}},\ and\
  \bibinfo {author} {\bibfnamefont {G.}~\bibnamefont {Mattoni}},\ }\bibfield
  {title} {\bibinfo {title} {Thirty years of puzzling superconductivity in
  sr$_2$ruo$_4$},\ }\href@noop {} {\bibfield  {journal} {\bibinfo  {journal}
  {Nature Physics}\ }\textbf {\bibinfo {volume} {20}},\ \bibinfo {pages} {1712}
  (\bibinfo {year} {2024})}\BibitemShut {NoStop}%
\bibitem [{\citenamefont {Puphal}\ \emph {et~al.}(2024)\citenamefont {Puphal},
  \citenamefont {Reiss}, \citenamefont {Enderlein}, \citenamefont {Wu},
  \citenamefont {Khaliullin}, \citenamefont {Sundaramurthy}, \citenamefont
  {Priessnitz}, \citenamefont {Knauft}, \citenamefont {Suthar}, \citenamefont
  {Richter}, \citenamefont {Isobe}, \citenamefont {van Aken}, \citenamefont
  {Takagi}, \citenamefont {Keimer}, \citenamefont {Suyolcu}, \citenamefont
  {Wehinger}, \citenamefont {Hansmann},\ and\ \citenamefont
  {Hepting}}]{puphal_unconventional_2024}%
  \BibitemOpen
  \bibfield  {author} {\bibinfo {author} {\bibfnamefont {P.}~\bibnamefont
  {Puphal}}, \bibinfo {author} {\bibfnamefont {P.}~\bibnamefont {Reiss}},
  \bibinfo {author} {\bibfnamefont {N.}~\bibnamefont {Enderlein}}, \bibinfo
  {author} {\bibfnamefont {Y.-M.}\ \bibnamefont {Wu}}, \bibinfo {author}
  {\bibfnamefont {G.}~\bibnamefont {Khaliullin}}, \bibinfo {author}
  {\bibfnamefont {V.}~\bibnamefont {Sundaramurthy}}, \bibinfo {author}
  {\bibfnamefont {T.}~\bibnamefont {Priessnitz}}, \bibinfo {author}
  {\bibfnamefont {M.}~\bibnamefont {Knauft}}, \bibinfo {author} {\bibfnamefont
  {A.}~\bibnamefont {Suthar}}, \bibinfo {author} {\bibfnamefont
  {L.}~\bibnamefont {Richter}}, \bibinfo {author} {\bibfnamefont
  {M.}~\bibnamefont {Isobe}}, \bibinfo {author} {\bibfnamefont {P.~A.}\
  \bibnamefont {van Aken}}, \bibinfo {author} {\bibfnamefont {H.}~\bibnamefont
  {Takagi}}, \bibinfo {author} {\bibfnamefont {B.}~\bibnamefont {Keimer}},
  \bibinfo {author} {\bibfnamefont {Y.~E.}\ \bibnamefont {Suyolcu}}, \bibinfo
  {author} {\bibfnamefont {B.}~\bibnamefont {Wehinger}}, \bibinfo {author}
  {\bibfnamefont {P.}~\bibnamefont {Hansmann}},\ and\ \bibinfo {author}
  {\bibfnamefont {M.}~\bibnamefont {Hepting}},\ }\bibfield  {title} {\bibinfo
  {title} {Unconventional crystal structure of the high-pressure superconductor
  la$_3$ni$_2$o$_7$},\ }\href@noop {} {\bibfield  {journal} {\bibinfo
  {journal} {Physical Review Letters}\ }\textbf {\bibinfo {volume} {133}},\
  \bibinfo {pages} {146002} (\bibinfo {year} {2024})}\BibitemShut {NoStop}%
\bibitem [{\citenamefont {Chen}\ \emph {et~al.}(2024)\citenamefont {Chen},
  \citenamefont {Zhang}, \citenamefont {Thind}, \citenamefont {Sharma},
  \citenamefont {LaBollita}, \citenamefont {Peterson}, \citenamefont {Zheng},
  \citenamefont {Phelan}, \citenamefont {Botana}, \citenamefont {Klie},\ and\
  \citenamefont {Mitchell}}]{chen_polymorphism_2024}%
  \BibitemOpen
  \bibfield  {author} {\bibinfo {author} {\bibfnamefont {X.}~\bibnamefont
  {Chen}}, \bibinfo {author} {\bibfnamefont {J.}~\bibnamefont {Zhang}},
  \bibinfo {author} {\bibfnamefont {A.~S.}\ \bibnamefont {Thind}}, \bibinfo
  {author} {\bibfnamefont {S.}~\bibnamefont {Sharma}}, \bibinfo {author}
  {\bibfnamefont {H.}~\bibnamefont {LaBollita}}, \bibinfo {author}
  {\bibfnamefont {G.}~\bibnamefont {Peterson}}, \bibinfo {author}
  {\bibfnamefont {H.}~\bibnamefont {Zheng}}, \bibinfo {author} {\bibfnamefont
  {D.~P.}\ \bibnamefont {Phelan}}, \bibinfo {author} {\bibfnamefont {A.~S.}\
  \bibnamefont {Botana}}, \bibinfo {author} {\bibfnamefont {R.~F.}\
  \bibnamefont {Klie}},\ and\ \bibinfo {author} {\bibfnamefont {J.~F.}\
  \bibnamefont {Mitchell}},\ }\bibfield  {title} {\bibinfo {title}
  {Polymorphism in the {Ruddlesden}-{Popper} {Nickelate}
  {La}$_{\textrm{3}}${Ni}$_{\textrm{2}}${O}$_{\textrm{7}}$: {Discovery} of a
  {Hidden} {Phase} with {Distinctive} {Layer} {Stacking}},\ }\href
  {https://doi.org/10.1021/jacs.3c14052} {\bibfield  {journal} {\bibinfo
  {journal} {Journal of the American Chemical Society}\ }\textbf {\bibinfo
  {volume} {146}},\ \bibinfo {pages} {3640} (\bibinfo {year}
  {2024})}\BibitemShut {NoStop}%
\bibitem [{\citenamefont {Wang}\ \emph {et~al.}(2024)\citenamefont {Wang},
  \citenamefont {Chen}, \citenamefont {Rutherford}, \citenamefont {Zhou},\ and\
  \citenamefont {Xie}}]{wang_long-range_2024}%
  \BibitemOpen
  \bibfield  {author} {\bibinfo {author} {\bibfnamefont {H.}~\bibnamefont
  {Wang}}, \bibinfo {author} {\bibfnamefont {L.}~\bibnamefont {Chen}}, \bibinfo
  {author} {\bibfnamefont {A.}~\bibnamefont {Rutherford}}, \bibinfo {author}
  {\bibfnamefont {H.}~\bibnamefont {Zhou}},\ and\ \bibinfo {author}
  {\bibfnamefont {W.}~\bibnamefont {Xie}},\ }\bibfield  {title} {\bibinfo
  {title} {Long-{Range} {Structural} {Order} in a {Hidden} {Phase} of
  {Ruddlesden}–{Popper} {Bilayer} {Nickelate}
  {La}$_{\textrm{3}}${Ni}$_{\textrm{2}}${O}$_{\textrm{7}}$},\ }\href
  {https://doi.org/10.1021/acs.inorgchem.3c04474} {\bibfield  {journal}
  {\bibinfo  {journal} {Inorganic Chemistry}\ }\textbf {\bibinfo {volume}
  {63}},\ \bibinfo {pages} {5020} (\bibinfo {year} {2024})}\BibitemShut
  {NoStop}%
\bibitem [{\citenamefont {Lin}\ and\ \citenamefont
  {Kulakov}(2004)}]{YBCO_detwin}%
  \BibitemOpen
  \bibfield  {author} {\bibinfo {author} {\bibfnamefont {C.}~\bibnamefont
  {Lin}}\ and\ \bibinfo {author} {\bibfnamefont {A.}~\bibnamefont {Kulakov}},\
  }\bibfield  {title} {\bibinfo {title} {In situ observation of ferroelastic
  detwinning of ybco single crystals by high temperature optical microscopy},\
  }\href {https://doi.org/https://doi.org/10.1016/j.physc.2004.02.022}
  {\bibfield  {journal} {\bibinfo  {journal} {Physica C: Superconductivity}\
  }\textbf {\bibinfo {volume} {408-410}},\ \bibinfo {pages} {27} (\bibinfo
  {year} {2004})},\ \bibinfo {note} {proceedings of the International
  Conference on Materials and Mechanisms of Superconductivity. High Temperature
  Superconductors VII -- M2SRIO}\BibitemShut {NoStop}%
\bibitem [{\citenamefont {Anderson}\ \emph {et~al.}(2024)\citenamefont
  {Anderson}, \citenamefont {Spai\ifmmode~\acute{c}\else \'{c}\fi{}},
  \citenamefont {Biniskos}, \citenamefont {Thompson}, \citenamefont {Yu},
  \citenamefont {Zwettler}, \citenamefont {Liu}, \citenamefont {Ye},
  \citenamefont {Granroth}, \citenamefont {Krogstad}, \citenamefont {Osborn},
  \citenamefont {Pelc},\ and\ \citenamefont {Greven}}]{Osborn2024}%
  \BibitemOpen
  \bibfield  {author} {\bibinfo {author} {\bibfnamefont {Z.~W.}\ \bibnamefont
  {Anderson}}, \bibinfo {author} {\bibfnamefont {M.}~\bibnamefont
  {Spai\ifmmode~\acute{c}\else \'{c}\fi{}}}, \bibinfo {author} {\bibfnamefont
  {N.}~\bibnamefont {Biniskos}}, \bibinfo {author} {\bibfnamefont
  {L.}~\bibnamefont {Thompson}}, \bibinfo {author} {\bibfnamefont
  {B.}~\bibnamefont {Yu}}, \bibinfo {author} {\bibfnamefont {J.}~\bibnamefont
  {Zwettler}}, \bibinfo {author} {\bibfnamefont {Y.}~\bibnamefont {Liu}},
  \bibinfo {author} {\bibfnamefont {F.}~\bibnamefont {Ye}}, \bibinfo {author}
  {\bibfnamefont {G.~E.}\ \bibnamefont {Granroth}}, \bibinfo {author}
  {\bibfnamefont {M.}~\bibnamefont {Krogstad}}, \bibinfo {author}
  {\bibfnamefont {R.}~\bibnamefont {Osborn}}, \bibinfo {author} {\bibfnamefont
  {D.}~\bibnamefont {Pelc}},\ and\ \bibinfo {author} {\bibfnamefont
  {M.}~\bibnamefont {Greven}},\ }\bibfield  {title} {\bibinfo {title}
  {Nanoscale structural correlations in a model cuprate superconductor},\
  }\href {https://doi.org/10.1103/PhysRevB.110.214519} {\bibfield  {journal}
  {\bibinfo  {journal} {Phys. Rev. B}\ }\textbf {\bibinfo {volume} {110}},\
  \bibinfo {pages} {214519} (\bibinfo {year} {2024})}\BibitemShut {NoStop}%
\bibitem [{\citenamefont {Bordet}\ \emph {et~al.}(1997)\citenamefont {Bordet},
  \citenamefont {Duc}, \citenamefont {Radaelli}, \citenamefont {Lanzara},
  \citenamefont {Saini}, \citenamefont {Bianconi},\ and\ \citenamefont
  {Antipov}}]{Hg1201}%
  \BibitemOpen
  \bibfield  {author} {\bibinfo {author} {\bibfnamefont {P.}~\bibnamefont
  {Bordet}}, \bibinfo {author} {\bibfnamefont {F.}~\bibnamefont {Duc}},
  \bibinfo {author} {\bibfnamefont {P.}~\bibnamefont {Radaelli}}, \bibinfo
  {author} {\bibfnamefont {A.}~\bibnamefont {Lanzara}}, \bibinfo {author}
  {\bibfnamefont {N.}~\bibnamefont {Saini}}, \bibinfo {author} {\bibfnamefont
  {A.}~\bibnamefont {Bianconi}},\ and\ \bibinfo {author} {\bibfnamefont
  {E.}~\bibnamefont {Antipov}},\ }\bibfield  {title} {\bibinfo {title}
  {Structural instability around tc observed in hg-1201 by neutron powder
  diffraction and exafs},\ }\href
  {https://doi.org/https://doi.org/10.1016/S0921-4534(97)90649-3} {\bibfield
  {journal} {\bibinfo  {journal} {Physica C: Superconductivity}\ }\textbf
  {\bibinfo {volume} {282-287}},\ \bibinfo {pages} {1081} (\bibinfo {year}
  {1997})},\ \bibinfo {note} {proceedings of the International Conference on
  Materials and Mechanisms of Superconductivity High Temperature
  Superconductors V Part II}\BibitemShut {NoStop}%
\bibitem [{\citenamefont {Salje}(1993)}]{Salje_book}%
  \BibitemOpen
  \bibfield  {author} {\bibinfo {author} {\bibfnamefont {E.~K.~H.}\
  \bibnamefont {Salje}},\ }\href@noop {} {\emph {\bibinfo {title} {Phase
  transitions in ferroelastic and co-elastic crystals}}},\ \bibinfo {edition}
  {student edition}\ ed.\ (\bibinfo  {publisher} {Cambridge University Press},\
  \bibinfo {address} {Cambridge, UK},\ \bibinfo {year} {1993})\BibitemShut
  {NoStop}%
\bibitem [{\citenamefont {Salje}(2010)}]{salje_multiferroic_2010}%
  \BibitemOpen
  \bibfield  {author} {\bibinfo {author} {\bibfnamefont {E.~K.~H.}\
  \bibnamefont {Salje}},\ }\bibfield  {title} {\bibinfo {title} {Multiferroic
  {Domain} {Boundaries} as {Active} {Memory} {Devices}: {Trajectories}
  {Towards} {Domain} {Boundary} {Engineering}},\ }\href@noop {} {\bibfield
  {journal} {\bibinfo  {journal} {ChemPhysChem}\ }\textbf {\bibinfo {volume}
  {11}},\ \bibinfo {pages} {940} (\bibinfo {year} {2010})}\BibitemShut
  {NoStop}%
\bibitem [{\citenamefont {Wright}(2025)}]{wright_fable-3dxrdimaged11_2025}%
  \BibitemOpen
  \bibfield  {author} {\bibinfo {author} {\bibfnamefont {J.~P.}\ \bibnamefont
  {Wright}},\ }\href {https://github.com/FABLE-3DXRD/ImageD11} {\bibinfo
  {title} {{FABLE}-{3DXRD}/{ImageD11}}} (\bibinfo {year} {2025})\BibitemShut
  {NoStop}%
\bibitem [{\citenamefont {Wright}\ \emph {et~al.}(2022)\citenamefont {Wright},
  \citenamefont {Giacobbe},\ and\ \citenamefont
  {Lawrence~Bright}}]{wright_using_2022}%
  \BibitemOpen
  \bibfield  {author} {\bibinfo {author} {\bibfnamefont {J.~P.}\ \bibnamefont
  {Wright}}, \bibinfo {author} {\bibfnamefont {C.}~\bibnamefont {Giacobbe}},\
  and\ \bibinfo {author} {\bibfnamefont {E.}~\bibnamefont {Lawrence~Bright}},\
  }\bibfield  {title} {\bibinfo {title} {Using {Powder} {Diffraction}
  {Patterns} to {Calibrate} the {Module} {Geometry} of a {Pixel} {Detector}},\
  }\href@noop {} {\bibfield  {journal} {\bibinfo  {journal} {Crystals}\
  }\textbf {\bibinfo {volume} {12}},\ \bibinfo {pages} {255} (\bibinfo {year}
  {2022})}\BibitemShut {NoStop}%
\bibitem [{\citenamefont {Bonnin}\ \emph {et~al.}(2014)\citenamefont {Bonnin},
  \citenamefont {Wright}, \citenamefont {Tucoulou},\ and\ \citenamefont
  {Palancher}}]{bonnin_impurity_2014}%
  \BibitemOpen
  \bibfield  {author} {\bibinfo {author} {\bibfnamefont {A.}~\bibnamefont
  {Bonnin}}, \bibinfo {author} {\bibfnamefont {J.~P.}\ \bibnamefont {Wright}},
  \bibinfo {author} {\bibfnamefont {R.}~\bibnamefont {Tucoulou}},\ and\
  \bibinfo {author} {\bibfnamefont {H.}~\bibnamefont {Palancher}},\ }\bibfield
  {title} {\bibinfo {title} {Impurity precipitation in atomized particles
  evidenced by nano x-ray diffraction computed tomography},\ }\href@noop {}
  {\bibfield  {journal} {\bibinfo  {journal} {Applied Physics Letters}\
  }\textbf {\bibinfo {volume} {105}},\ \bibinfo {pages} {084103} (\bibinfo
  {year} {2014})}\BibitemShut {NoStop}%
\bibitem [{\citenamefont {van Aarle}\ \emph {et~al.}(2016)\citenamefont {van
  Aarle}, \citenamefont {Palenstijn}, \citenamefont {Cant}, \citenamefont
  {Janssens}, \citenamefont {Bleichrodt}, \citenamefont {Dabravolski},
  \citenamefont {De~Beenhouwer}, \citenamefont {Joost~Batenburg},\ and\
  \citenamefont {Sijbers}}]{van_aarle_fast_2016}%
  \BibitemOpen
  \bibfield  {author} {\bibinfo {author} {\bibfnamefont {W.}~\bibnamefont {van
  Aarle}}, \bibinfo {author} {\bibfnamefont {W.~J.}\ \bibnamefont
  {Palenstijn}}, \bibinfo {author} {\bibfnamefont {J.}~\bibnamefont {Cant}},
  \bibinfo {author} {\bibfnamefont {E.}~\bibnamefont {Janssens}}, \bibinfo
  {author} {\bibfnamefont {F.}~\bibnamefont {Bleichrodt}}, \bibinfo {author}
  {\bibfnamefont {A.}~\bibnamefont {Dabravolski}}, \bibinfo {author}
  {\bibfnamefont {J.}~\bibnamefont {De~Beenhouwer}}, \bibinfo {author}
  {\bibfnamefont {K.}~\bibnamefont {Joost~Batenburg}},\ and\ \bibinfo {author}
  {\bibfnamefont {J.}~\bibnamefont {Sijbers}},\ }\bibfield  {title} {\bibinfo
  {title} {Fast and flexible {X}-ray tomography using the {ASTRA} toolbox},\
  }\href@noop {} {\bibfield  {journal} {\bibinfo  {journal} {Optics Express}\
  }\textbf {\bibinfo {volume} {24}},\ \bibinfo {pages} {25129} (\bibinfo {year}
  {2016})}\BibitemShut {NoStop}%
\bibitem [{\citenamefont {Zhang}\ \emph {et~al.}(2019)\citenamefont {Zhang},
  \citenamefont {Lun}, \citenamefont {Li},\ and\ \citenamefont
  {Zhou}}]{zhang_method_2019}%
  \BibitemOpen
  \bibfield  {author} {\bibinfo {author} {\bibfnamefont {Y.}~\bibnamefont
  {Zhang}}, \bibinfo {author} {\bibfnamefont {M.~C.}\ \bibnamefont {Lun}},
  \bibinfo {author} {\bibfnamefont {C.}~\bibnamefont {Li}},\ and\ \bibinfo
  {author} {\bibfnamefont {Z.}~\bibnamefont {Zhou}},\ }\bibfield  {title}
  {\bibinfo {title} {Method for improving the spatial resolution of narrow
  x-ray beam-based x-ray luminescence computed tomography imaging},\
  }\href@noop {} {\bibfield  {journal} {\bibinfo  {journal} {Journal of
  Biomedical Optics}\ }\textbf {\bibinfo {volume} {24}},\ \bibinfo {pages}
  {086002} (\bibinfo {year} {2019})}\BibitemShut {NoStop}%
\bibitem [{\citenamefont {Kim}\ \emph {et~al.}(2023)\citenamefont {Kim},
  \citenamefont {Hayashi},\ and\ \citenamefont {Yabashi}}]{kim_emergence_2023}%
  \BibitemOpen
  \bibfield  {author} {\bibinfo {author} {\bibfnamefont {J.}~\bibnamefont
  {Kim}}, \bibinfo {author} {\bibfnamefont {Y.}~\bibnamefont {Hayashi}},\ and\
  \bibinfo {author} {\bibfnamefont {M.}~\bibnamefont {Yabashi}},\ }\bibfield
  {title} {\bibinfo {title} {The emergence of super-resolution beyond the probe
  size in scanning {3DXRD} microscopy},\ }\href@noop {} {\bibfield  {journal}
  {\bibinfo  {journal} {Journal of Synchrotron Radiation}\ }\textbf {\bibinfo
  {volume} {30}},\ \bibinfo {pages} {1108} (\bibinfo {year}
  {2023})}\BibitemShut {NoStop}%
\end{thebibliography}%

\clearpage
\newpage

\onecolumngrid

\section*{SUPPLEMENTAL MATERIAL}
\setcounter{page}{1}
\setcounter{figure}{0}
\setcounter{table}{0}
\setcounter{section}{0}
\renewcommand{\thepage}{S\arabic{page}}
\renewcommand{\thesection}{S\arabic{section}}
\renewcommand{\thetable}{S\arabic{table}}
\renewcommand{\thefigure}{S\arabic{figure}}
\newcounter{SIfig}
\renewcommand{\theSIfig}{S\arabic{SIfig}}

\section{3DXRD measurements}
Single crystal samples were selected from the same growth batch as in \cite{tidey_pronounced_2022}. A single crystal of approximate dimensions 15~$\times$~20~$\times$~20 \textmu m was secured on a MiTeGen cryoloop using a UV-curing resin. This prevents the crystal from moving during data collection. 3DXRD measurements were performed at the ID11 beamline at the ESRF using a beam energy of 43.5 keV. The beam was focused to 120 nm in size. Sinogram scans were taken as the sample was alternately rotated forwards and backwards over 180$^\circ$ with a step size of 0.05$^\circ$. The diffractometer was translated with a step size of 50 nm over a 26 \textmu m range. Scans took around 2 hours. Measurements were taken at 300, 140, 120 and 100 K. All measurements were performed at the same nominal position in $z$, though slight variations may result from thermal contraction of the sample holder with cooling. 

\clearpage

\section{Data processing}
All data were processed using the ImageD11 software package \cite{wright_fable-3dxrdimaged11_2025}. The raw sinogram data were obtained at each temperature. Individual diffraction spots were identified, corrected \cite{wright_using_2022} and converted to reciprocal space coordinates using instrument calibration from a silicon single crystal. Peaks were indexed using an averaged orthorhombic unit cell for 300 and 140 K and a tetragonal unit cell for 120 and 100 K. ($hkl$) indices were then assigned to each observed reflection. 

Reconstruction of the crystal shape, as shown in Fig.~\ref{1}b, proceeded using tomographic back projection of the peak intensities versus diffractometer $y$-coordinate. The reconstruction was enhanced using the Maximum Likelihood Expectation Maximisation (MLEM) algorithm \cite{bonnin_impurity_2014}. Local variations in the lattice parameters within the crystal were fit using an adaptation of the tomographic algorithm recently described by Henningsson and Hall \cite{henningsson_efficient_2023} and implemented using the ASTRA toolbox \cite{van_aarle_fast_2016}. The local lattice parameters within each voxel were refined by fitting the observed peak positions to the metric tensor equation from powder diffraction:
\begin{equation}
    \frac{1}{d^{2}}=Ah^{2}+Bk^{2}+Cl^{2}+Dkl+Ehl+Fhk
\end{equation}

An initially uniform 2D image of the metric tensor elements ($A-F$) was forward projected to generate a sinogram by weighting each of the six layers according to the grain shape reconstruction. The resulting $1/d^2$ values for each $hkl$ projection in the 2D sinogram were then compared to observed peak positions. The fit was optimised using a linear least-squares method (L-BFGS algorithm in SciPy) to minimise the sum of squared differences between calculated and observed $1/d^2$ values. This refinement procedure applied no symmetry constraints or regularisation, enabling an unbiased reconstruction of local lattice distortions. Unlike conventional strain tensor approaches, this method does not require an explicit reference `d-zero' lattice, as the refinement directly fits the metric tensor elements.

The refined metric tensor elements were used to calculate the local lattice basis vectors for each voxel. These were then used to compute symmetry-adapted strain modes with respect to the $I4/mmm$ aristotype structure, where $a = b \neq c$ and $\alpha = \beta = \gamma = 90^\circ$. Reference values of $a$ and $c$ were defined as the average lattice parameters across all voxels at each temperature. 

\clearpage

\section{Symmetry-adapted strain analysis}

To compute the symmetry-adapted strain modes, the refined lattice basis vectors were first expressed in Cartesian coordinates and compared to the basis vectors of the parent cell to obtain unit-less parent cell strains ($P_1-P_6$). The parent is adopted in the $F4/mmm$ setting, which shares the same basis as LTO/LTT, to avoid a later strain transformation to the distorted setting. Unit-less parent cells strains ($P_1-P_6$) were obtained according to the following relationships:

\begin{align}
P_1 &= \frac{a(x)}{a_0(x)} - 1 \\[10pt] \label{p1}
P_2 &= \frac{b(y) - b_0(x) \cdot \frac{P_6}{2}}{b_0(y)} - 1 \\[10pt]
P_3 &= \frac{1}{c_0(z)} \left( c(z) - c_0(x) \cdot \frac{P_5}{2} - c_0(y) \cdot \frac{P_4}{2} \right) - 1 \\[10pt]
P_4 &= \frac{2}{c_0(z)} \left( c(y) - c_0(x) \cdot \frac{P_6}{2} - c_0(y) \cdot (P_2 + 1) \right) \\[10pt]
P_5 &= \frac{2}{c_0(z)} \left( c(x) - c_0(x) \cdot \frac{a(x)}{a_0(x)} - c_0(y) \cdot \frac{P_6}{2} \right) \\[10pt]
P_6 &= \frac{2}{b_0(y)} \left( b(x) - b_0(x) \cdot (P_1 + 1) \right) \label{p6}
\end{align}

where $a(x,y,z),~b(x,y,z),~a(x,y,z)$ denote the Cartesian components of the distorted lattice basis vectors, and $a_0(x,y,z),~b_0(x,y,z),~c_0(x,y,z)$ denote the corresponding components of the parent lattice basis vectors. 

The symmetry-adapted strain modes relative to the $I4/mmm$ aristotype were computed as:
\begin{align}
&\Gamma_1^+(1) = \frac{P_1 + P_2}{\sqrt{2}} \\[10pt] \label{g1}
&\Gamma_1^+(2) = P_3 \\[10pt]
&\Gamma_2^+ = \frac{P_6}{\sqrt{2}} \\[10pt]
&\Gamma_4^+ = \frac{P_1 - P_2}{\sqrt{2}} \\[10pt]
&\Gamma_5^+(a) = \frac{P_4}{\sqrt{2}} \\[10pt]
&\Gamma_5^+(b) = \frac{P_5}{\sqrt{2}} \label{g5b}
\end{align}

\begin{figure}[!h]
\includegraphics[width=0.6\textwidth]{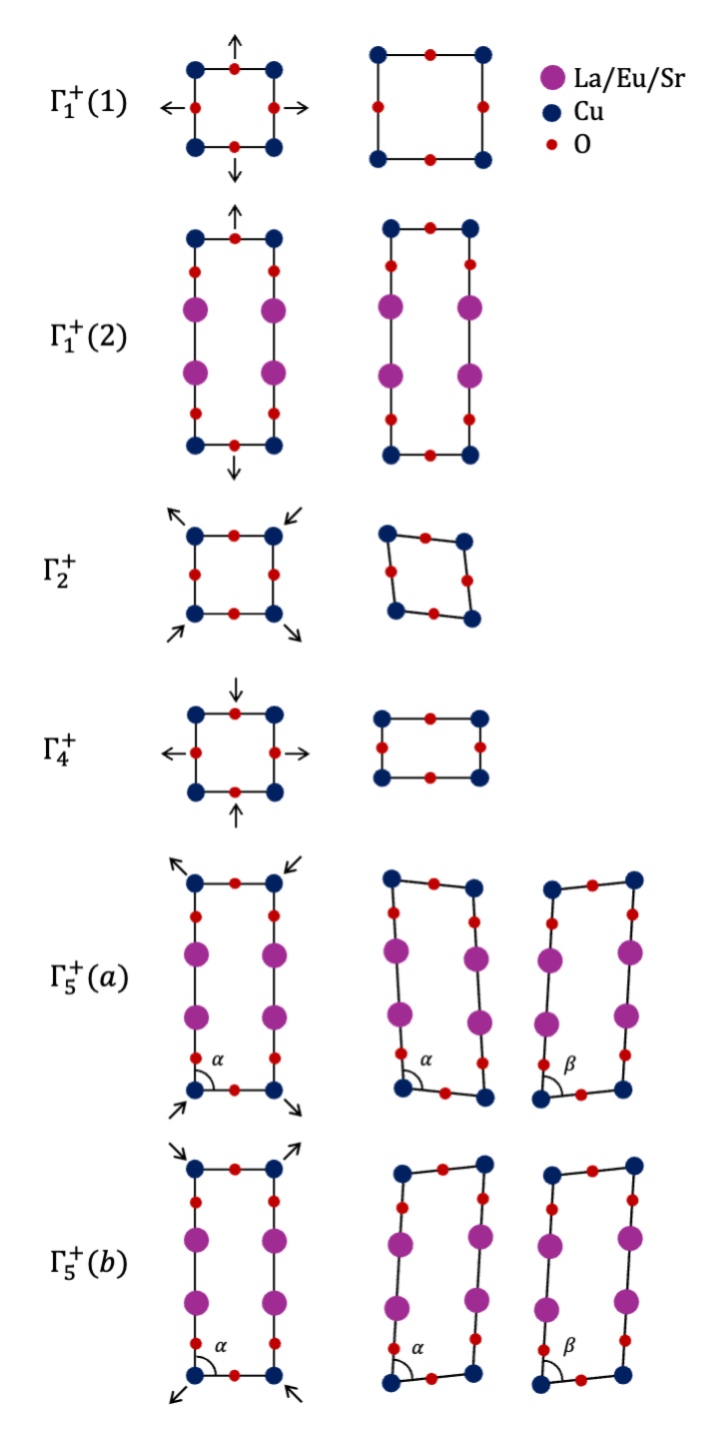}
\caption{Schematic of strain modes, shown relative to the LTO/LTT basis.}
\label{s1}
\end{figure}

Each mode corresponds to a distortion relative to the high symmetry $I4/mmm$ aristotype. $\Gamma_{1}^{+}(1)$  corresponds to a symmetric expansion or contraction within the ab plane which preserves tetragonal symmetry. $\Gamma_{1}^{+}(2)$ describes a uniform elongation or compression along the $c$ axis. $\Gamma_{2}^{+}$ is a shear strain in the $ab$ plane. $\Gamma_{4}^{+}$ represents an orthorhombic strain, corresponding to an antisymmetric distortion between the $a$ and $b$ axes (i.e., a $\neq$ b). Finally, $\Gamma_{5}^{+}(a)$ and $\Gamma_{5}^{+}(b)$ describe shear strains between the $ab$-plane and the $c$ axis. Fig.~\ref{s1} schematically depicts these strains.

\clearpage

\section{Full strain modes distributions for 100 K, 120 K and 140 K}

\begin{figure}[!h]
\includegraphics[width=\textwidth]{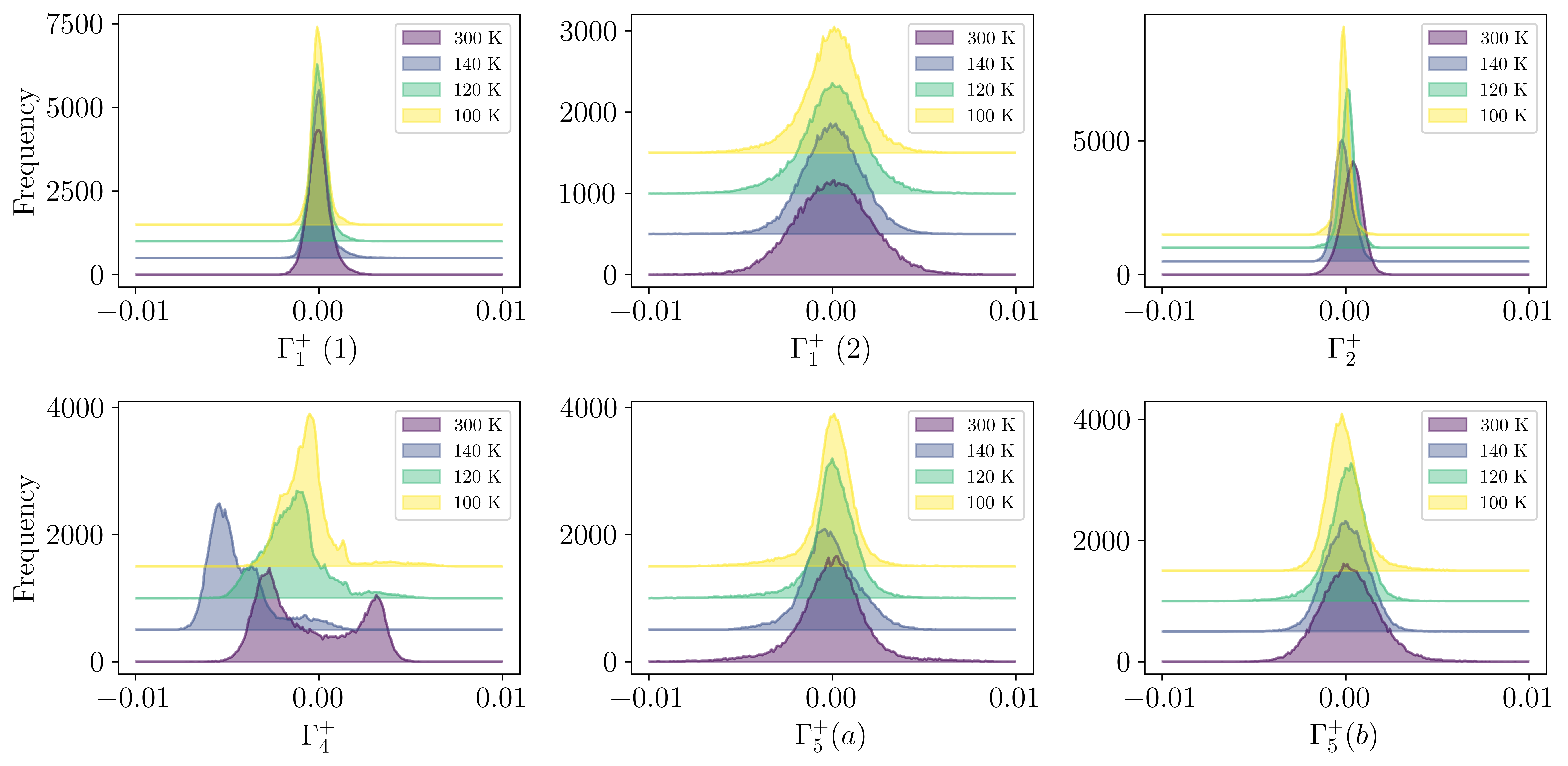}
\caption{Histograms of symmetry-adapted strain modes for all temperatures.}
\label{s2}
\end{figure}

\begin{figure}[!h]
\includegraphics[width=\textwidth]{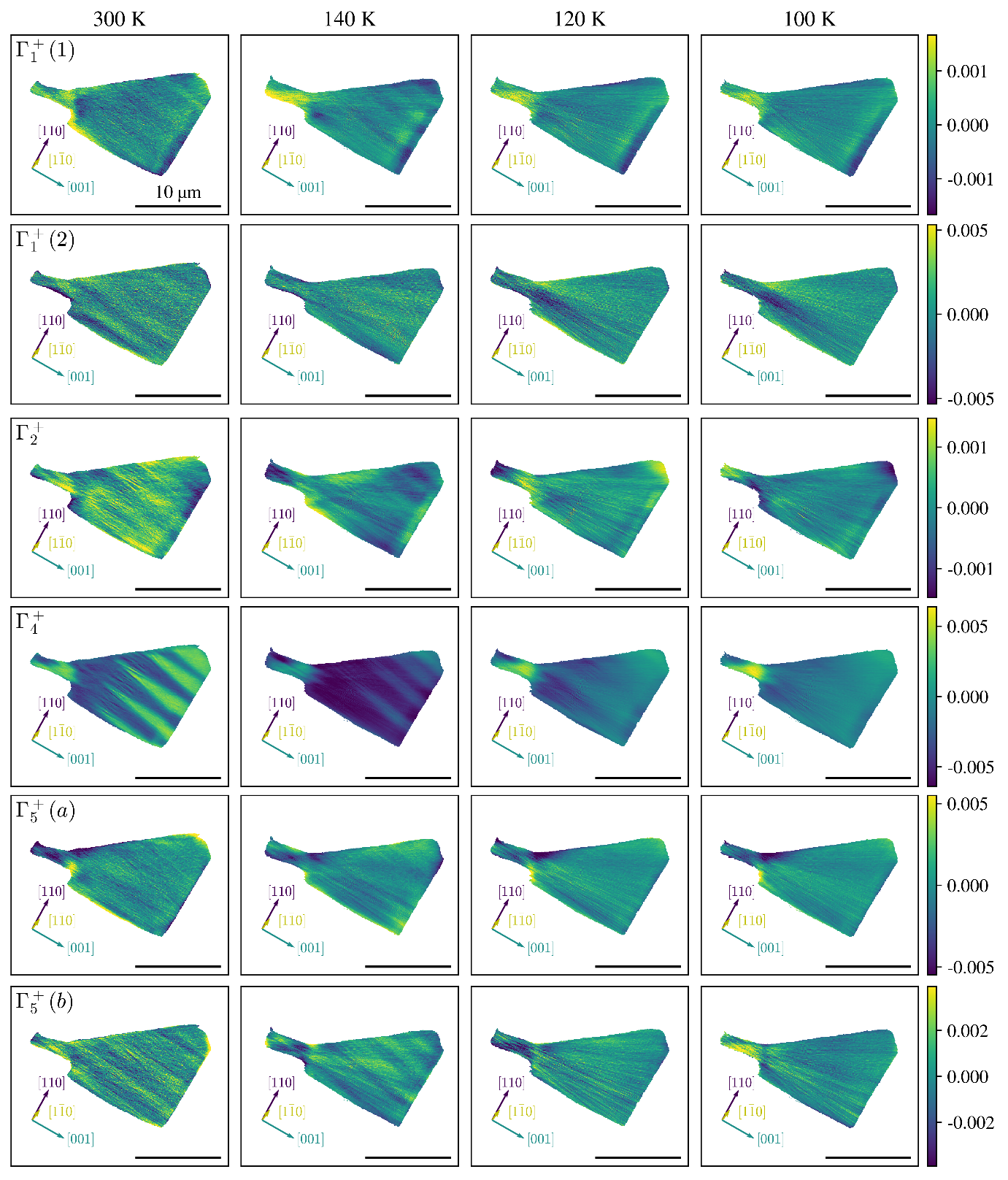}
\caption{Spatially resolved maps of symmetry-adapted strain modes at all temperatures. }
\label{s3}
\end{figure}

\begin{figure}[!h]
\includegraphics[width=\textwidth]{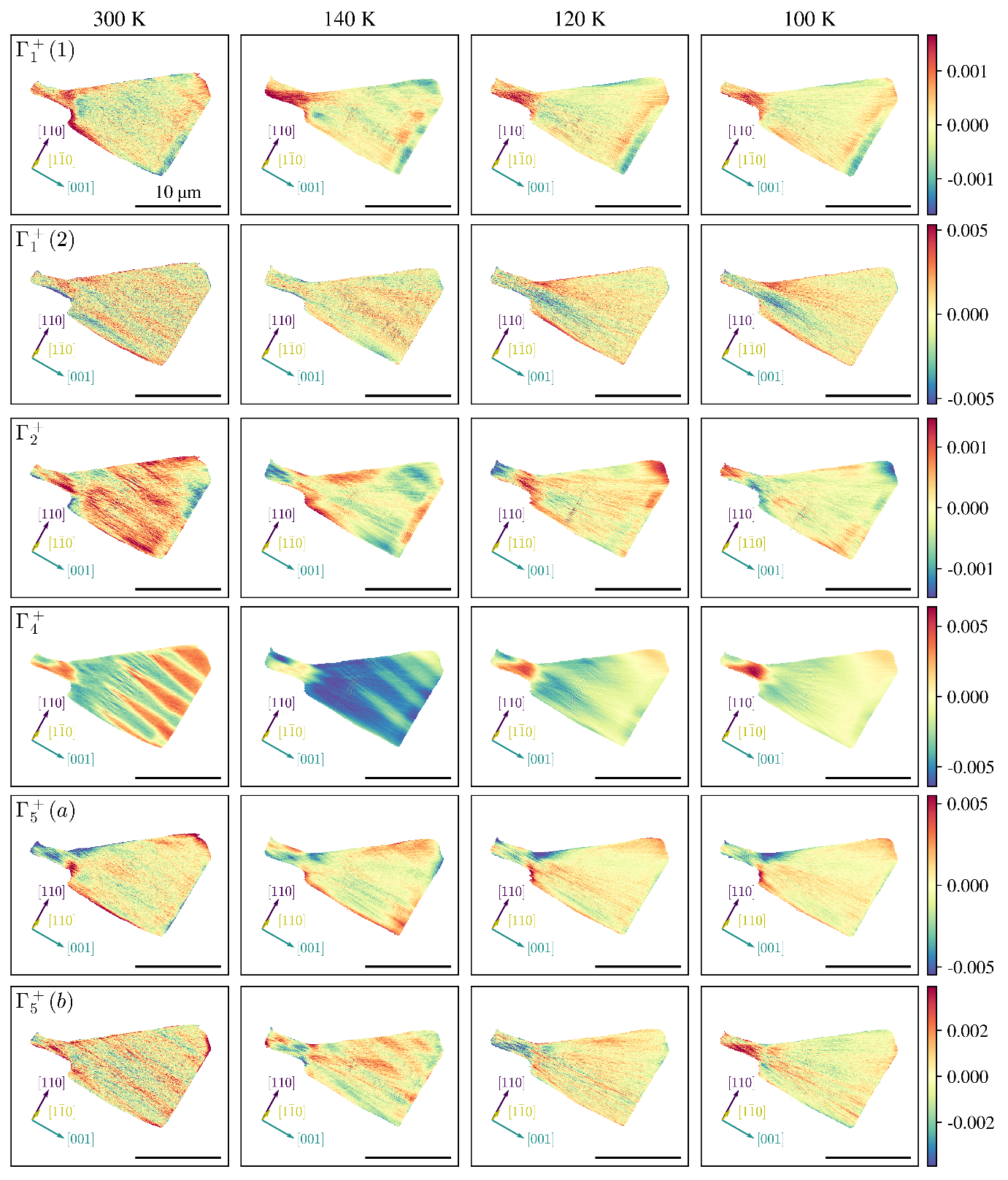}
\caption{Spatially resolved maps of symmetry-adapted strain modes at all temperatures with diverging colour map.}
\label{s4}
\end{figure}

\clearpage

\section{Microscopic/crystallographic construction of domain walls}

$Pccn$ is characterised by octahedral tilts about [110] and [1$\bar{1}$0] (equivalent to [100] and [010] for the HTT cell), with alternating layers tilting in opposite senses. It resembles the LTT phase, but the tilts about the two axes differ in magnitude (and are usually smaller), transforming as X$_{3}^{+}(a;b)$ relative to HTT.

\begin{figure}[!h]
\includegraphics[width=\textwidth]{domain_wall.png}
\caption{Representation of domain wall with LTT symmetry at centre. $A$-site cation omitted for clarity. Oxygens are shown in teal and magenta indicating whether they are above and below the CuO$_2$ planes, respectively.}
\label{dw}
\end{figure}

\clearpage

\section{Further details for line profiles for LESCO at 300 K}

\begin{figure}[!h]
\includegraphics[width=0.7\textwidth]{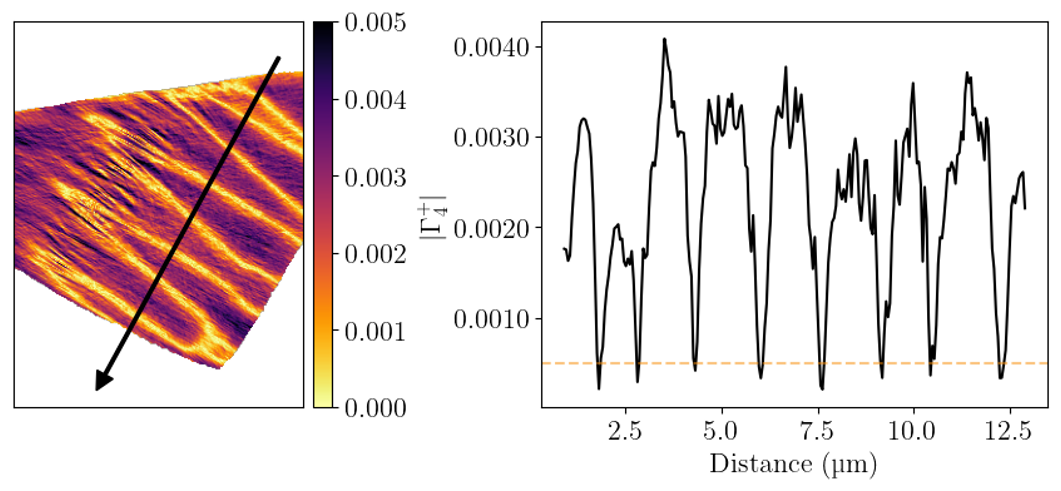}
\caption{Line profile of LESCO at 300 K. }
\label{s6}
\end{figure}

\begin{table}[!h]
  \setlength{\tabcolsep}{6.0pt}
  \caption{Line profile peak widths for LESCO at 300 K, corresponding to Figure S6.}
  \label{tabs1}
  \centering
  \begin{tabular}{|c|c|}
    \hline  \hline \rule{0pt}{1.2\normalbaselineskip}

    Peak & Width at 4$~\times~10^{-4}$ \\ \hline

    1 & 0.12\\ \hline
    2 & 0.08 \\ \hline
    3 & 0.08  \\ \hline
    4 & 0.13  \\ \hline
    5 & 0.14 \\ \hline
    6 & 0.09  \\ \hline
    7 & 0.05  \\ \hline
    8 & 0.17  \\ 

    \hline 
    \hline
\end{tabular}
\end{table}

\clearpage

\section{Resolution of measurements}

We find our spatial resolution to be consistently between 30 -- 70 nm at a strain-resolution of 2$~\times~10^{-4}$. The strain resolution of ID11 in the setup used for the present experiment is better than 2$~\times~10^{-4}$. Additionally, studies on scanning-based imaging techniques have shown that oversampling can enhance the spatial resolution beyond the limit set by the beam size \cite{zhang_method_2019, kim_emergence_2023}. 

\begin{figure}[!h]
\includegraphics[width=0.5\textwidth]{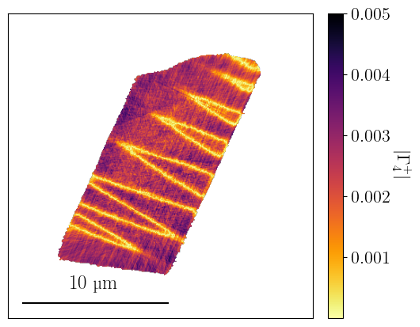}
\caption{$|\Gamma_{4}^{+}|$ strain map for CST.}
\label{s7}
\end{figure}

\begin{figure}[!h]
\includegraphics[width=0.7\textwidth]{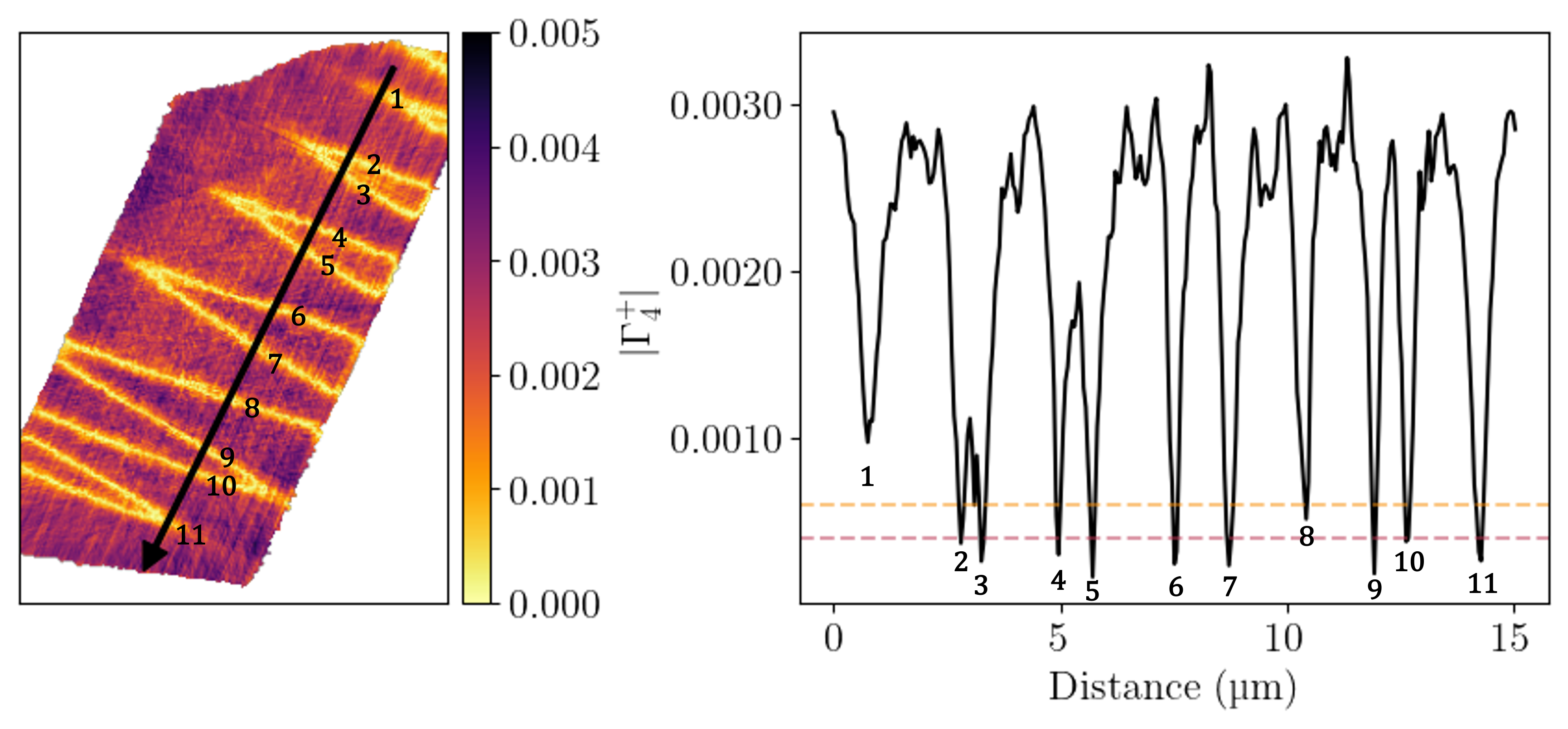}
\caption{$|\Gamma_{4}^{+}|$ strain map and line profile for CST.}
\label{s8}
\end{figure}

The intrinsic spatial resolution of the reconstructions in our experiment is established by studying a ferroic material, $n = 2$ Ruddlesden-Popper Ca$_{2.15}$Sr$_{0.85}$Ti$_{2}$O$_{7}$ (CST), where the existence of giant domain walls have previously been evidenced in the related system Ca$_{3}$Ti$_{2}$O$_{7}$ \cite{smith_infrared_2019}. At 300 K, CST crystallises in the orthorhombic space group $A2_1am$, derived from the same tetragonal aristotype structure ($I4/mmm$) as LESCO. It exhibits a comparable orthorhombic ferroelastic domain structure, characterised by tetragonal-like domain walls, as shown in Figs.~\ref{s7}-\ref{s8}. 

\begin{figure}[!h]
\includegraphics[width=0.7\textwidth]{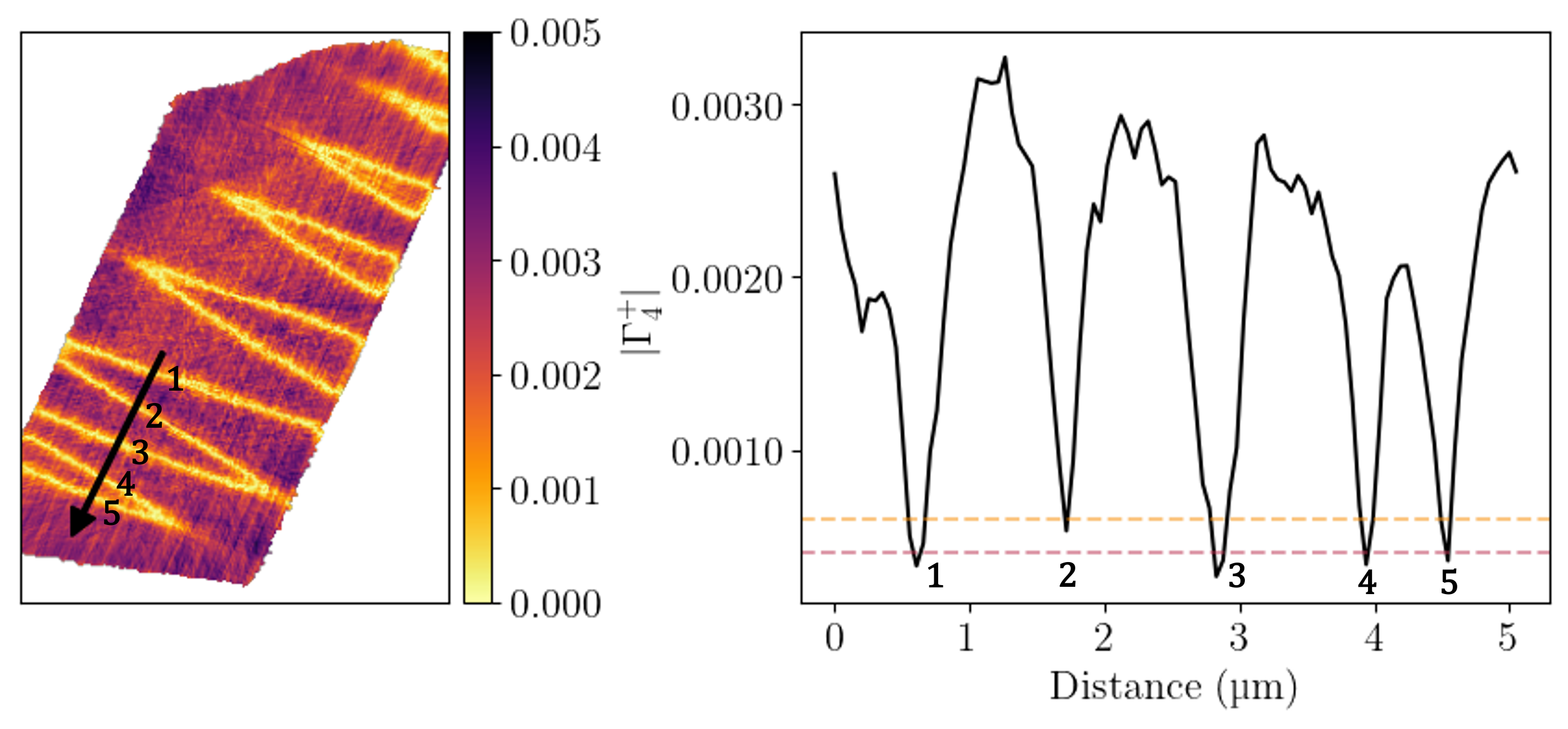}
\caption{$|\Gamma_{4}^{+}|$ strain map and line profile for CST.}
\label{s9}
\end{figure}

\begin{figure}[!h]
\includegraphics[width=\textwidth]{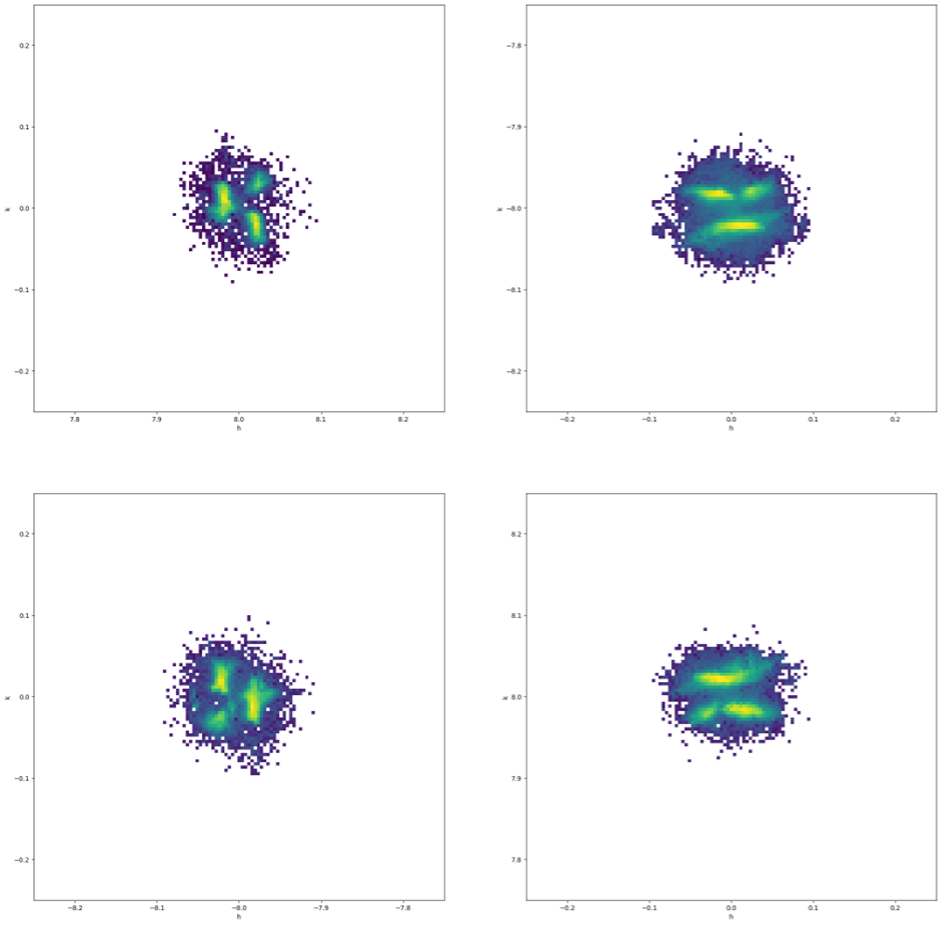}
\caption{Reciprocal space reconstruction generated from raw diffraction data showing that three distinct domains (two LTO and one LTT) are clearly resolved.}
\label{s10}
\end{figure}

The domain walls in CST appear slightly sharper than those observed in LESCO, providing an internal check on the spatial resolution of the 3DXRD measurements. Since CST was measured in the same experiment, using the same beam size and translation step, any difference in wall sharpness reflects real structural behaviour rather than instrumental factors. Line profiles from different regions of the crystal, presented in Figs.~\ref{s9}-\ref{s10}, highlight these features in more detail.

If the centre of the domain wall is tetragonal, $|\Gamma_{4}^{+}|$ should be zero at this point. In practice, our measurements record a minimum of approximately 2$~\times~10^{-4}$, which reflects the effective strain resolution of the experiment. The spatial resolution is determined from this from the average change in strain from 4$~\times~10^{-4}$ to 6$~\times~10^{-4}$ across several domain walls. From this average change, the spatial resolution is determined to be 61 nm.

\begin{table}[ht!]
  \setlength{\tabcolsep}{6.0pt}
  \caption{Line profile peak widths for CST at 300 K, corresponding to Figure S8.}
  \label{tabs2}
  \centering
  \begin{tabular}{|c|c|c|c|}
    \hline  \hline \rule{0pt}{1.2\normalbaselineskip}

    Peak & Width at 4$~\times~10^{-4}$ & Width at 6$~\times~10^{-4}$ & Difference \\ \hline

    1 & - & - & - \\ \hline
    2 & - & - & - \\ \hline
    3 & 0.07 & 0.13 & 0.06 \\ \hline
    4 & 0.04 & 0.10 & 0.06 \\ \hline
    5 & 0.06 & 0.12 & 0.06 \\ \hline
    6 & 0.07 & 0.11 & 0.04 \\ \hline
    7 & 0.09 & 0.17 & 0.08 \\ \hline
    8 & - & 0.04 & - \\ \hline
    9 & 0.05 & 0.09 & 0.04 \\ \hline
    10 & 0.05 & 0.11 & 0.06 \\ \hline
    11 & 0.09 & 0.17 & 0.08 \\ 

    \hline 
    \hline
\end{tabular}
\end{table}

\begin{table}[ht!]
  \setlength{\tabcolsep}{6.0pt}
  \caption{Line profile peak widths for CST at 300 K, corresponding to Figure S9.}
  \label{tabs3}
  \centering
  \begin{tabular}{|c|c|c|c|}
    \hline  \hline \rule{0pt}{1.2\normalbaselineskip}

    Peak & Width at 4$~\times~10^{-4}$ & Width at 6$~\times~10^{-4}$ & Difference \\ \hline

    1 & 0.05 & 0.12 & 0.07 \\ \hline
    2 & - & 0.02 & - \\ \hline
    3 & 0.07 & 0.12 & 0.05 \\ \hline
    4 & 0.02 & 0.09 & 0.07 \\ \hline
    5 & 0.01 & 0.07 & 0.06 \\ 

    \hline 
    \hline
\end{tabular}
\end{table}

\clearpage

Fig.~\ref{s10} shows the reciprocal space reconstruction generated from the raw diffraction data, where diffraction spots that must originate from three distinct domains (two LTO and the LTT domains) are clearly resolved.

\end{document}